\documentclass[twocolumn]{aastex6}
\usepackage{amssymb, amsmath}
\newcommand{\Lx}{L_{\rm x}}
\newcommand{\Lrot}{L_{\rm rot}}
\newcommand{\Frot}{F_{\rm rot}}
\newcommand{\ymod}{y_{\rm model}}
\newcommand{\Fx}{F_{\rm x}}
\newcommand{\eq}[1]{\label{#1} }              
\newcommand{\exx}[2]{#1 \times 10^{#2}}       
\renewcommand{\vec}[1]{\mbox{\boldmath $#1$}} 

\begin{document}
\title{
X-ray and Rotational Luminosity Correlation
and Magnetic Heating of the Radio Pulsars}
\shorttitle{
X-ray and Rotational Luminosity Correlation
}

\author{S. Shibata\altaffilmark{1}
E. Watanabe\altaffilmark{2}
Y. Yatsu\altaffilmark{3}
T. Enoto\altaffilmark{4, 5}
and 
A. Bamba\altaffilmark{6, 7, 8}
}
\altaffiltext{1}{
Department of Physics, Yamagata University, 
1-4-12 Kojirakawa, Yamagata, JAPAN}
\altaffiltext{2}{
School of Science and Technology, Yamagata University,  
1-4-12 Kojirakawa, Yamagata, JAPAN}
\altaffiltext{3}{
Department of Physics, Tokyo Institute of Technology,
2-12-1 Ohokayama, Meguro, Tokyo 152-8551, Japan}
\altaffiltext{4}{
The Hakubi Center for Advanced Research, Kyoto University, Kyoto 606-8302, Japan}
\altaffiltext{5}{
Department of Astronomy, Kyoto University, Kitashirakawa-Oiwake-cho, Sakyo-ku, Kyoto 606-8502, Japan}
\altaffiltext{6}{
Department of Physics and Mathematics, Aoyama Gakuin University,
5-10-1 Fuchinobe, Chuo-ku, Sagamihara, Kanagawa 252-5258, Japan}
\altaffiltext{7}{
Department of Physics, The University of Tokyo, 7-3-1 Hongo,
Bunkyo-ku, Tokyo 113-0033, Japan}
\altaffiltext{8}{
Research Center for the Early Universe, School of Science, The
University of Tokyo, 7-3-1 Hongo, Bunkyo-ku, Tokyo 113-0033, Japan}

\begin{abstract}
Previous works have suggested a correlation
between the X-ray luminosity $\Lx$ and the rotational luminosity $\Lrot$
of the radio pulsars.
However,
none of the obtained regression lines are statistically acceptable
due to large  scatters.

We construct a statistical model which has
an intrinsic $\Lx$--$\Lrot$ relation and reproduces
the observed $\Lx$ distribution about it
by using a Monte Carlo simulator,
which takes into account the effects 
obscuring the intrinsic relation,
i.e., 
the anisotropy of radiation, additional heating, 
uncertainty in distance and detection limit of the instruments.

From the ATNF pulsar catalog
we collect 57 `ordinary radio pulsars' 
with significant detection and 42 with upper limits.
The sample does not include the high-magnetic field pulsars ($>10^{13}$~G),
which are separately analyzed.

We obtain a statistically acceptable relation 
$\Lx (0.5 - 10 {\rm keV})= 10^{31.69} (\Lrot / L_0)^{c_1}$ with
$c_1 = 1.03 \pm 0.27$ and $L_0 =10^{35.38}$.
The distribution about the obtained $\Lx$--$\Lrot$ relation is
reproduced well by the simulator.

Pulsars with abnormally high $\Lx$ fall into two types:
one is the soft gamma-ray pulsars, and the other is
thermally bright pulsars 
in comparison with the standard cooling curve.
On the other hand,
pulsars showing low $\Lx$ are found to  have dim pulsar wind nebulae.
We argue that there is an unknown mechanism that governs
both the magnetospheric emission and the pulsar wind nebulae,
and it might involve
the production rate of electron-positron pairs.

The high-field pulsars form a distinctive population 
other than the ordinary pulsars due to their excess luminosities.

\end{abstract}

\keywords{pulsars; general -- stars: neutron -- X-rays: general}

\section{Introduction}

Empirical relation between the high-energy luminosity
and the rotational luminosity provides a good
constraint on the 
emission and particle acceleration 
mechanisms of the pulsar magnetosphere.
The rotational luminosity is given by $\Lrot = 4 \pi I \dot{P} / P^3$
where  $P$ is the observed pulse period, 
$\dot{P}$ is its time derivative and
$I$ is the moment of inertia of the neutron star,
which is assume to be $\approx 10^{45}$g~cm$^2$.
(The standard deviation of the neutron star mass is $\sim 0.2$
solar masses, which is $\sim 0.1$ in log-scale
(Ozel et al. 2015, Kiziltan et al. 2013),
so the constancy of $I$ would not
affect the present statistical study.
The proper motion, $\sim 450$~km~s$^{-1}$, is also unimportant
in determining $\Lrot$.) 
{\it Fermi} LAT observations have indicated that 
the gamma-ray luminosities $L_\gamma$ follow
$L_\gamma \propto \Lrot^{1/2}$ for young pulsars
(Abdo et al. 2013),
while the X-ray luminosities show a steeper correlation
(Becker \& Truemper 1997), which
may be roughly represented by $\Lx \sim 10^{-3} \Lrot$.
The slope of one half  implies that 
the gamma-rays originate from the primary particles
the flux of which is proportional to the Goldreich-Julian 
current $\sim (\Lrot /c)^{1/2}$.
On the other hand, a steeper slope in $\Lx - \Lrot$ relation may be
due to the fact that secondary pairs are attributed to  X-ray radiation.
A further implication on the $L_\gamma - \Lrot$ relation is
the change of the slope from one half to about unity when 
$\Lrot$ is decreased (Marelli et al. 2011).
This indicates that some qualitative change in the particle acceleration
and/or emission mechanisms 
when the available voltage is reduced to $\sim 10^{13}$ Volt.
Takata et al.(2011) show that the outer gap model accounts for this change. 

Even though the recent observations from  the X-ray satellites
such as {\it Chandra} and {\it XMM-Newton}
provides high-quality spectral data for significant numbers of objects,
the $\Lx - \Lrot$ correlation is still quite uncertain.
The first identification
of the correlation $\Lx = 10^{-16.8} \Lrot^{1.39}$
was suggested by Seward and Wang (1988).
Later several suggestions have been made;
$\Lx = 10^{-3} \Lrot$ 
(Becker and Trueumper, 1997),
$\Lx = 10^{-12} \Lrot^{1.5}$ 
(Saito 1998),
$\Lx = 10^{-15.34} \Lrot^{1.34}$ 
(Possenti et al. 2002),
$\Lx = 10^{-0.8} \Lrot^{0.92}$ 
(Li et al. 2008),
$\Lx = 10^{-3.24} \Lrot^{0.997}$ in the  0.1--2~keV,
and
$\Lx = 10^{-15.72} \Lrot^{1.336}$ in the  2--10~keV 
(Becker 2009).
The difference in the slopes may be due to the choice
of energy bands and
of the components: 
thermal, non thermal and pulsar wind nebula (PWN).
All these works suggest correlations 
with linear regressions. 
Nevertheless, none of the regression lines are statistically acceptable, 
i.e., scatter about the regression lines are  significant
(Possenti et al. 2002, Kargaltsev et al. 2012).

\begin{figure}
\begin{center}
\includegraphics[width=9cm]{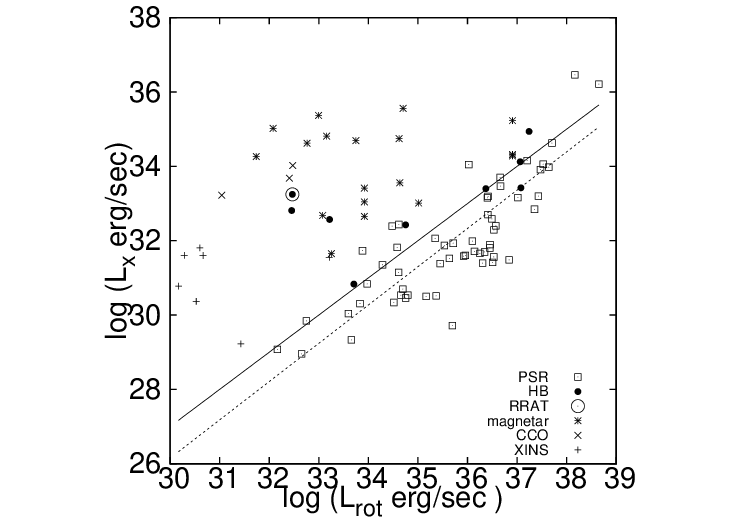}
\caption{ \label{LxLrot1}
The $\Lx - \Lrot$ plot for our sample
and related objects.
The open squares and the filled circles are, respectively,
for Sample SAB (ordinary radio pulsars) and Sample HB
(high-magnetic field pulsars) defined in section \ref{sample}.
The data for other neutron stars,
the magnetars  (Olausen \& Kaspi 2014), 
CCO (Halpern \& Gotthelf 2010), and
XINS (Vigan{\`o} et al. 2013)
are superposed. 
The large open circle indicates RRAT.
The dashed line is the best fit model relation obtained in section \ref{result}
while the solid line is the earlier suggestion $\Lx = 10^{-3} \Lrot $
(Becker \& Truemper, 1997).
}
\end{center}
\end{figure}

Figure~1 shows the  $\Lx - \Lrot$ plot for  our sample,
which is described in detail in \S \ref{sample}. 
The open squares indicate the ordinary radio pulsars. 
It indicates  a correlation between $\Lx$ and $\Lrot$.
However, a large scatter can also be seen.
As suggested by Kargaltsev et al. (2012),
the scatter is too large to be explained by 
incorrectly determined distances.
Some pulsars appear to be much dimmer than
the regression lines suggested earlier
the solid line in Figure~\ref{LxLrot1}).
This may be a geometrical effect, with which
the phase averaged flux tends to be smaller than the true flux
depending on viewing angles: the observed flux
can be very small if the viewing angle is bad.
Some authors suggest a critical line below which 
all the data locate, i.e.,
$\Lx < 10^{-18.5} \Lrot^{1.48}$
(Possenti et al. 2002), and 
$\Lx < 10^{-21.4} \Lrot^{1.51}$
(Kargaltsev et al., 2012).
Because large scatter is also seen in the luminosity of PWN,
these authors have suggested that some unknown physics (or other effect)
restrains the X-ray luminosity.
We will discuss this point 
in section~\ref{discussion}.
Observational detection limits could
also affect the $\Lx - \Lrot$ plot;
a dim object may not be observed if its distance is large.
This selection effect brings about
less distribution below an expected correlation
when $\Lrot$ is small.

We take into account the above-mentioned effects in a Monte Carlo
simulator and compare the simulated distribution with the observed data,
and thereby
we attempt to find a statistically acceptable 
`hidden' relation between $\Lx$ and $\Lrot$.
We also statistically-test the distribution 
about the $\Lx - \Lrot$ relation.
This test enables us to search for characteristic
of the scatter.

Another reason the $\Lx - \Lrot$ correlation 
may be contaminated is the X-ray radiation which is 
not caused by the rotational energy and
therefore not related to $\Lrot$.
There are noticeable neutron star populations other than
the rotation powered pulsars; more specifically they are the magnetars 
which is a joint population of anomalous X-ray pulsar (AXP) and
soft gamma-ray repeater (SGR), 
the central compact object (CCO),
the X-ray isolated neutron star (XINS). 
All these objects show excess X-ray luminosity.
The origin is supposed to be the neutron star cooling radiation and 
dissipation of magnetic field. 

Let us briefly summarize the properties of these X-ray sources.
Persistent luminosity of magnetars taken from the McGill Magnetar Catalogue
(Olausen \& Kaspi, 2014) are included in Figure~1.
The main characteristics of a proto-typical magnetars are
(1) persistent X-ray luminosity in the range $10^{33}-10^{36}$erg~s$^{-1}$, 
which exceeds $\Lrot$, 
(2) high time-variability and 
(3) large breaking torque,
which implies the surface dipole field of $B_d \sim 10^{14} - 10^{15}$~G
(see e.g., Turolla et al. 2015 for a review), where
the dipole field is derived from $P$ and $\dot{P}$
by
\begin{eqnarray}
B_d & = &
\left(
{3 \over 2} {I c^3 \over (2 \pi)^2 R_*^6 } P \dot{P}
\right)^{1/2}
\\
& \approx &
 1.1 \times 10^{12}{\rm G} \left( {P \over 1 {\rm s}} \; 
{ \dot{P} \over 10^{-15} } \right)^{1/2} ,
\end{eqnarray}
where $R_* \approx 10^6$~cm is the radius of the neutron star.
By definition,
the high X-ray luminosity and bursting activity of the magnetars
are powered by their strong magnetic fields
(Duncan \& Thompson, 1992).
However, it was found recently that
SGR~0418$+$5729 has a dipole field of $B_d \sim 6.1 \times 10^{12}$~G
(Esposito et al., 2010; Rea et al., 2010; Rea et al., 2013),
which is well inside the range of ordinary radio pulsars.
This fact suggests that 
the magnetic field of magnetars is in multipole components $B_m$, which
can be larger than and independent of $B_d$.
The origin of $B_m$ may be a large toroidal field in the crust.
In the theoretical point of view,
Ciolfi \& Rezzolla (2013) shows that
the toroidal field can be much stronger than the poloidal field.

CCOs are bright X-ray sources which reside near the centers of SNRs.
The spin down parameters of three CCOs (PSR~J1852+0040 in Kes 79,
PSR~0821--4300 in Puppis A and 1E1207.4-5209 in PKS 1209-51/52)
are measured
(Halpern \& Gotthelf 2010, Gotthelf et al. 2013).
It is suggested that
ages of CCOs are much larger than those of SNRs, and
they  have small magnetic field, $B_d \sim 10^{10}$~G
(e.g., Bogdanov et al. 2014).
The X-ray luminosity of these three CCOs are also plotted in Figure~1.
Their positions in the $\Lx - \Lrot$ plot are obviously apart from the
general trend of the radio pulsars and rather in the end of the magnetar group.
Although $B_d$ is small, presently-`hidden' or past strong crust field
may exist.
All the known CCOs exhibit no radio emission.
However, their locations in $P - \dot{P}$ diagram are within the
range of the ordinary radio pulsars.
This suggests that
the ordinary radio pulsars associated with SNR
may have dissipative crustal field 
similar to CCOs.
This possibility was examined for nearby objects ($< 6$~kpc) 
by Bogdanov et al. (2014), who found no
X-ray excess emission in their sample.

Several tens of radio pulsars are known to have dipole fields
larger than $\sim 10^{13}$G.
PSR~J1718-3718  and PSR~J1734-3333, respectively, have 
$B_d = 7.47 \times 10^{13}$~G and $B_d =5.23 \times 10^{13 }$~G,
which are comparable with $5.9 \times 10^{13}$~G
of the AXP 1E~2259+586.
At present, three of the high-magnetic field pulsars exhibit
large X-ray luminosity ($> 0.1 \; \Lrot$).
If $B_m$ is independent of $B_d$,
ordinary radio pulsars may have dissipative crustal field.
such as those seen in magnetars and X-ray excess luminosity. 
Occurrence rate of such excess may depend on $B_d$ or 
on the evolutionary path in
$P - \dot{P}$ diagram. 
Another important fact on the high-magnetic field pulsar is that
PSR~J1846--0258 with $B_d = \exx{5}{13}$~G was thought to be a rotation 
powered pulsar (though radio quiet) 
but showed X-ray outbursts in a way commonly seen 
in the magnetars (Gavriil et al. 2008).
Quite recently, the radio pulsar J1119--6127 with 
$B_d \sim \exx{4.1}{13}$~G exhibited a magnetar-like
outburst, adding the second example
(Younes et al. 2016;
Kennea et al. 2016;
Archibald et al. 2016).
This suggests that some of ordinary radio pulsars may have 
dissipative magnetic fields, which are not always in the dipole field
but may be in crustal multipole fields.
If dissipation is sudden, it causes outbursts, while gradual dissipation would
cause an excess of the persistent X-ray luminosity or
a high surface temperature as compared with standard cooling curves of the neutron star.
If this is the case, magnetic heating could cause scatter in the $\Lx - \Lrot$ plot
of the ordinary radio pulsars. 

In this paper,
we investigate statistical properties of the $\Lx - \Lrot$ plot,
searching for an intrinsic $\Lx - \Lrot$ relation.
We discuss the difference between the ordinary radio pulsars and 
the high-magnetic field pulsars. We also discuss the origin of the scatter: 
why some pulsars show considerably large $\Lx$ while
some show very small $\Lx$.

The paper is organized as follows:
observational data are accumulated and the statistical samples are provided
in \S~\ref{sample}, 
the method of statistical analysis is given in \S~\ref{method},
and
the result is given in \S~\ref{result} and discussed
in \S~\ref{discussion}.

\section{Sample Preparation} \label{sample}

To define our sample, we use the ATNF pulsar catalogue\footnote{Available at
http://www.atnf.csiro.au/research/pulsar/psrcat/.}
(Manchester et al. 2005), 
and make a set of `ordinary' radio pulsars satisfying following conditions:
(1) its $P$, $\dot{P}$ and the distance $d$ are available,
(2) it is observed in radio,
(3) not a magnetar; neither AXP nor SGR\footnote{%
The objects listed in McGill magnetar catalog
(Olausen \& Kaspi 2014;
http://www.physics.mcgill.ca/\~pulsar/magnetar/main.html)
or indicated by AXP in the ATNF catalogue are removed.},
(4) not a millisecond pulsar (MSP) (filtered by $B_d > 10^{10}$G),
(to exclude a possible effect of history of accretion),
(5) not in a binary system.
Since the catalog provides the best estimate distance rather than
the dispersion measure distance if it is available, we use the best estimate
distance in the catalog.
There are two main reasons we exclude MSPs in the present
sample. The weak field of MSP is though to be due to mass accretion, so that
`buried' field might exists and causes additional heating
(Bejger et al., 2011).
From the observational point of view, the study of the second Fermi catalog
(Abdo et al., 2013)
suggests that gamma-X ratios of the ordinary pulsar and MSP are different. 

After the filtering, 
we list a high-magnetic field sample out of the obtained set
with the condition $B_d > 10^{13}$~G to define `Sample HB'.
The remaining pulsars are the `ordinary' radio pulsars and are separated
into five groups according the `rotational energy flux'
defined by $\Frot = \Lrot / 4 \pi d^2$,
i.e.,
`Sample S' ($\Frot > 10^{-9}$~erg/cm$^2$s$^{-1}$),
`Sample A' ($10^{-9} \geq \Frot > 10^{-10}$~erg/cm$^2$s$^{-1}$),
`Sample B' ($10^{-10} \geq \Frot > 10^{-11}$~erg/cm$^2$s$^{-1}$),
`Sample C' ($10^{-11} \geq \Frot > 10^{-12}$~erg/cm$^2$s$^{-1}$)  and
`Sample D' ($10^{-12} \geq \Frot $~erg/cm$^2$s$^{-1}$).
Since the lowest value 
of $\Frot$ for Sample B is 
$10^{-11}$~erg/cm$^2$s$^{-1}$,
its expected flux would be 
$\approx 10^{-14}$~erg/cm$^2$s$^{-1}$, which is roughly the detection
limit of the current instruments with typical exposure times.
This means that still dimmer Samples C and D are not useful 
in determining the $\Lx - \Lrot$ relation.
In our statistical analysis, we use 
the joint Sample S$+$A$+$B (simply denoted by `Sample SAB')
and Sample HB. To see selection effect, we also use Sample S, A and B
separately.

\begin{figure}
\begin{center}
\includegraphics[width=10cm]{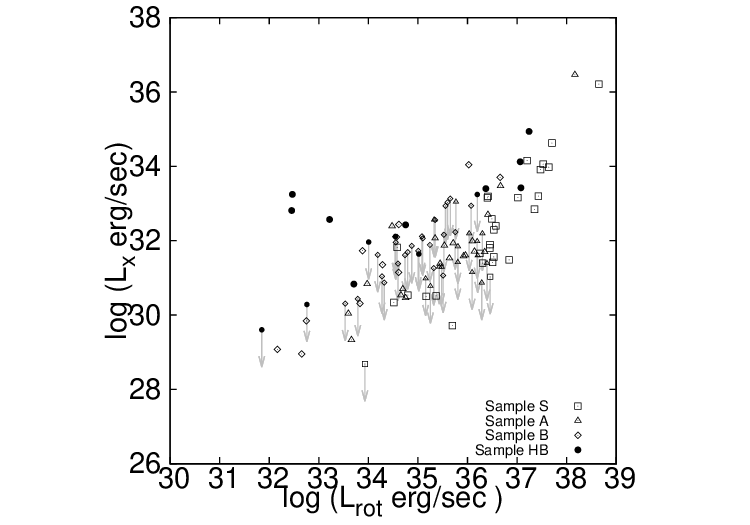}
\caption{ \label{LxLrot2}
The $\Lx - \Lrot$ plot for our sample
in Table~\ref{dset}. 
The upper limit values are indicated by the arrows.
}
\end{center}
\end{figure}

We review the data of the listed pulsars in the literature.
Using the reported models for the observed unabsorbed flux,
we convert them into the values in the 0.5--10~keV band.
As has been mentioned in the previous section,
we are interested in the possibility of magnetic heating 
in the ordinary radio pulsars, so that the choice is the 0.5--10~keV band
rather than 2--10~keV.
As for the most of the detected data,
we use the flux corrected for the presence of a nebula component,
if available.
We use the distances cited in the ATNF pulsar catalogue to obtain the luminosity
(not the values in the original papers).
Since uncertainty in the distance causes scattering in $\Lx$,
this effect is taken into account the probability density distribution
obtained by the Monte Carlo simulator.
In our statistical analysis, we also use the upper limit when they are
available.
Apart from reviewing the past publications, 
we searched HEASARC archive data for
all the pulsars in Sample S, A, B and HB.
Only for PSR~J1909$+$0749, there is no published data, but
we find a HEASARC archive data for serendipitous pointing toward
the object and find an upper limit of 
$\Lx < 9.648 \times 10^{-14}$~erg/cm$^2$s$^{-1}$
 in the $0.5-10$ keV band
assuming the power law model with a photon index of 1.5
(detail is given in Appendix B).
There are 61 objects in Sample SAB and 41 objects in Sample HB,
for which no observation is made or 
the obtained upper limit is too large and useless ($> 10^{-2} \Lrot$).
The sample names have subscripts `d' or `ul' to indicate 
if detected values or upper limits, i.e.,
Sample S$_{\rm d}$,
Sample S$_{\rm ul}$, 
Sample A$_{\rm d}$, 
Sample A$_{\rm ul}$ and so on.
The numbers of samples for each subset are summarized in Table~\ref{nosample}.
The data sets are given in Table~\ref{dset} and shown in Figure~\ref{LxLrot2}.

\begin{table*}
\caption{Summary of the Numbers of our Samples. \label{nosample}}
\begin{center}
\begin{tabular}{ccrrrr} \hline \hline
Sample Name &  Range & total & detected &  upper limit & not observed \\
\hline
S & $\Frot > 10^{-9}$                & 29 & 27 & 2  & 0 \\
A & $10^{-9} \geq \Frot > 10^{-10}$  & 43 & 20 & 15 & 8 \\
B & $10^{-10} \geq \Frot > 10^{-11}$ & 88 & 10 & 25 & 53 \\
SAB & $\Frot > 10^{-11}$             & 160 & 57 & 42 & 61 \\
HB& $B_d > 10^{13}$                  & 56 &  9 &  6 & 41 \\
\hline
\end{tabular}
\end{center}
\end{table*}

\begin{longtable*}{lrrrrrr}
\caption{$\Lx - \Lrot$ data for all the samples \label{dset} 
} \\ \hline 
Name & Period & distance & $\log \Lrot$ & $\log \Lx$  & $\log (\Lx /\Lrot )$ & References \\
PSR & s & kpc & erg~s$^{-1}$ & erg~s$^{-1}$ &  & \\ \hline
\endfirsthead
\caption[]{(continued)} \\ \hline
Name & Period & distance & $\log \Lrot$ & $\log \Lx$  & $\log (\Lx / \Lrot )$ & References \\
PSR & s & kpc & erg~s$^{-1}$ & erg~s$^{-1}$ &  & \\ \hline
\endhead
\hline  \multicolumn{7}{l}{Sample S$_{\rm d}$ }  \\
B0531+21 &   0.033 &   2.00 &   38.650 &  36.209 &    -2.44&  Kargaltsev \& Pavlov 2008 \\ 
B0656+14 &   0.385 &   0.28 &   34.581 &  31.821 &    -2.76&  De Luca et al., 2005 \\ 
B0833-45 &   0.089 &   0.28 &   36.840 &  31.483 &    -5.36&  Kargaltsev \& Pavlov 2008 \\ 
B0906-49 &   0.107 &   1.00 &   35.693 &  29.715 &    -5.98&  Kargaltsev et al. 2012 \\ 
B1046-58 &   0.124 &   2.90 &   36.303 &  31.396 &    -4.91&  Kargaltsev \& Pavlov 2008 \\ 
B1706-44 &   0.102 &   2.60 &   36.533 &  32.289 &    -4.24&  Kargaltsev \& Pavlov 2008 \\ 
B1757-24 &   0.125 &   4.61 &   36.413 &  33.188 &    -3.23&  Kargaltsev \& Pavlov 2008 \\ 
B1823-13 &   0.101 &   4.12 &   36.454 &  31.895 &    -4.56&  Kargaltsev \& Pavlov 2008 \\ 
B1951+32 &   0.040 &   3.00 &   36.572 &  32.400 &    -4.17&  Kargaltsev \& Pavlov 2008 \\ 
B2334+61 &   0.495 &   0.70 &   34.797 &  30.537 &    -4.26&  McGowan et al., 2006 \\ 
J0205+6449 &   0.066 &   3.20 &   37.431 &  33.201 &    -4.23&  Kargaltsev \& Pavlov 2008 \\ 
J0633+1746 &   0.237 &   0.25 &   34.513 &  30.337 &    -4.18&  Kargaltsev \& Pavlov 2008 \\ 
J1357-6429 &   0.166 &   4.09 &   36.491 &  32.588 &    -3.90&  Kargaltsev \& Pavlov 2008 \\ 
J1400-6325 &   0.031 &   7.00 &   37.705 &  34.628 &    -3.08&  Renaud et al., 2010 \\ 
J1420-6048 &   0.068 &   7.65 &   37.015 &  33.157 &    -3.86&  Kargaltsev \& Pavlov 2008 \\ 
J1524-5625 &   0.078 &   3.84 &   36.508 &  31.418 &    -5.09&  Kargaltsev et al. 2012 \\ 
J1617-5055 &   0.069 &   6.46 &   37.203 &  34.152 &    -3.05&  Kargaltsev \& Pavlov 2008 \\ 
J1732-3131 &   0.197 &   0.80 &   35.163 &  30.500 &    -4.66&  Ray et al., 2011, \\ 
J1740+1000 &   0.154 &   1.36 &   35.365 &  30.515 &    -4.85&  Kargaltsev \& Pavlov 2008 \\ 
J1747-2809 &   0.052 &  17.55 &   37.638 &  33.977 &    -3.66&  Porquet et al., 2003 \\ 
J1747-2958 &   0.099 &   2.49 &   36.399 &  33.152 &    -3.25&  Kargaltsev \& Pavlov 2008 \\ 
J1809-1917 &   0.083 &   3.71 &   36.249 &  31.659 &    -4.59&  Kargaltsev \& Pavlov 2008 \\ 
J1833-1034 &   0.062 &   4.10 &   37.527 &  34.057 &    -3.47&  Kargaltsev \& Pavlov 2008 \\ 
J1907+0602 &   0.107 &   3.01 &   36.451 &  31.804 &    -4.65&  Abdo et al., 2010 \\ 
J2021+3651 &   0.104 &   1.80 &   36.529 &  31.562 &    -4.97&  Kargaltsev \& Pavlov 2008 \\ 
J2022+3842 &   0.049 &  10.00 &   37.472 &  33.913 &    -3.56&  Marthi et el., 2011 \\ 
J2229+6114 &   0.052 &   3.00 &   37.351 &  32.845 &    -4.51&  Kargaltsev \& Pavlov 2008 \\ 

\hline  \multicolumn{7}{l}{Sample S$_{\rm ul}$ }  \\
B1742-30 &   0.367 &   0.20 &  33.930 & $<$ 28.683 & $<$  -5.25  &  Prinz \& Becker 2015 \\ 
J1913+1011 &   0.036 &   4.48 &  36.458 & $<$ 31.029 & $<$  -5.43  &  Prinz \& Becker 2015 \\ 

\hline  \multicolumn{7}{l}{Sample A$_{\rm d}$  }  \\
B0114+58 &   0.101 &   2.14 &   35.345 &  32.063 &    -3.28&  Prinz \& Becker 2015 \\ 
B0355+54 &   0.156 &   1.00 &   34.657 &  30.533 &    -4.12&  Kargaltsev \& Pavlov 2008 \\ 
B0540-69 &   0.050 &  53.70 &   38.167 &  36.462 &    -1.70&  Kaaret et al., 2001 \\ 
B1055-52 &   0.197 &   1.53 &   34.478 &  32.392 &    -2.09&  De Luca et al., 2005 \\ 
B1338-62 &   0.193 &   8.55 &   36.141 &  31.710 &    -4.43&  Prinz \& Becker 2015 \\ 
B1800-21 &   0.134 &   4.40 &   36.345 &  31.695 &    -4.65&  Kargaltsev \& Pavlov 2008 \\ 
B1822-09 &   0.769 &   0.30 &   33.659 &  29.331 &    -4.33&  Prinz \& Becker 2015 \\ 
B1853+01 &   0.267 &   3.30 &   35.633 &  31.525 &    -4.11&  Kargaltsev \& Pavlov 2008 \\ 
B1929+10 &   0.227 &   0.31 &   33.595 &  30.041 &    -3.55&  Kargaltsev \& Pavlov 2008 \\ 
J0538+2817 &   0.143 &   1.30 &   34.694 &  30.698 &    -4.00&  Kargaltsev \& Pavlov 2008 \\ 
J0729-1448 &   0.252 &   4.37 &   35.447 &  31.381 &    -4.07&  Kargaltsev et al. 2012 \\ 
J1016-5857 &   0.107 &   9.31 &   36.411 &  32.699 &    -3.71&  Kargaltsev \& Pavlov 2008 \\ 
J1028-5819 &   0.091 &   2.76 &   35.920 &  31.585 &    -4.34&  Kargaltsev et al. 2012 \\ 
J1509-5850 &   0.089 &   3.85 &   35.712 &  31.930 &    -3.78&  Kargaltsev \& Pavlov 2008 \\ 
J1531-5610 &   0.084 &   3.10 &   35.957 &  31.605 &    -4.35&  Kargaltsev et al. 2012 \\ 
J1702-4128 &   0.182 &   5.18 &   35.534 &  31.870 &    -3.66&  Kargaltsev et al. 2012 \\ 
J1718-3825 &   0.075 &   4.24 &   36.097 &  31.988 &    -4.11&  Kargaltsev et al. 2012 \\ 
J1741-2054 &   0.414 &   0.30 &   33.977 &  30.837 &    -3.14&  Camilo et al., 2009 \\ 
J1856+0245 &   0.081 &  10.29 &   36.665 &  33.469 &    -3.20&  Rousseau et al., 2012 \\ 
J2043+2740 &   0.096 &   1.13 &   34.752 &  30.456 &    -4.29&  Abdo et al., 2013 \\ 

\hline  \multicolumn{7}{l}{Sample A$_{\rm ul}$ }  \\
B0740-28 &   0.167 &   2.00 &  35.155 & $<$ 30.979 & $<$  -4.18  &  Prinz \& Becker 2015 \\ 
B1727-33 &   0.139 &   4.26 &  36.091 & $<$ 31.146 & $<$  -4.95  &  Prinz \& Becker 2015 \\ 
B1830-08 &   0.085 &   4.50 &  35.766 & $<$ 33.045 & $<$  -2.72  &  Kargaltsev et al., 2012 \\ 
J0248+6021 &   0.217 &   2.00 &  35.328 & $<$ 32.564 & $<$  -2.76  &  Abdo et al., 2013 \\ 
J0940-5428 &   0.088 &   4.27 &  36.287 & $<$ 30.870 & $<$  -5.42  &  Prinz \& Becker 2015 \\ 
J1105-6107 &   0.063 &   7.07 &  36.393 & $<$ 31.390 & $<$  -5.00  &  Prinz \& Becker 2015 \\ 
J1637-4642 &   0.154 &   5.77 &  35.806 & $<$ 31.845 & $<$  -3.96  &  Prinz \& Becker 2015 \\ 
J1702-4310 &   0.241 &   5.44 &  35.803 & $<$ 31.417 & $<$  -4.39  &  Prinz \& Becker 2015 \\ 
J1739-3023 &   0.114 &   3.41 &  35.478 & $<$ 31.291 & $<$  -4.19  &  Prinz \& Becker 2015 \\ 
J1828-1101 &   0.072 &   7.26 &  36.194 & $<$ 31.983 & $<$  -4.21  &  Prinz \& Becker 2015 \\ 
J1831-0952 &   0.067 &   4.33 &  36.033 & $<$ 32.191 & $<$  -3.84  &  Prinz \& Becker 2015 \\ 
J1835-1106 &   0.166 &   3.08 &  35.250 & $<$ 30.777 & $<$  -4.47  &  Prinz \& Becker 2015 \\ 
J1837-0604 &   0.096 &   6.19 &  36.301 & $<$ 32.193 & $<$  -4.11  &  Prinz \& Becker 2015 \\ 
J1841-0345 &   0.204 &   4.15 &  35.430 & $<$ 31.297 & $<$  -4.13  &  Prinz \& Becker 2015 \\ 
J1928+1746 &   0.069 &   8.12 &  36.206 & $<$ 31.599 & $<$  -4.61  &  Prinz \& Becker 2015 \\ 

\hline  \multicolumn{7}{l}{Sample B$_{\rm d}$  }  \\
B0540+23 &   0.246 &   3.54 &   34.611 &  31.148 &    -3.46&  Prinz \& Becker 2015 \\ 
B0628-28 &   1.244 &   0.32 &   32.164 &  29.078 &    -3.09&  Tepedelenlio{\u{g}}lu \& {\"{O}}gelman 2005 \\ 
B0823+26 &   0.531 &   0.32 &   32.655 &  28.952 &    -3.70&  Becker et al., 2004 \\ 
B0919+06 &   0.431 &   1.10 &   33.831 &  30.309 &    -3.52&  Prinz \& Becker 2015 \\ 
B0950+08 &   0.253 &   0.26 &   32.748 &  29.845 &    -2.90&  Becker et al., 2004 \\ 
B1221-63 &   0.216 &   4.00 &   34.285 &  31.351 &    -2.93&  Prinz \& Becker 2015 \\ 
B1822-14 &   0.279 &   5.45 &   34.615 &  32.436 &    -2.18&  Bogdanov et al., 2014 \\ 
J0855-4644 &   0.065 &   9.90 &   36.025 &  34.044 &    -1.98&  Acero et al., 2013 \\ 
J1112-6103 &   0.065 &  30.00 &   36.657 &  33.703 &    -2.95&  Prinz \& Becker 2015 \\ 
J1301-6310 &   0.664 &   2.06 &   33.881 &  31.729 &    -2.15&  Prinz \& Becker 2015 \\ 

\hline  \multicolumn{7}{l}{Sample B$_{\rm ul}$ }  \\
B0136+57 &   0.272 &   2.60 &  34.320 & $<$ 30.869 & $<$  -3.45  &  Prinz \& Becker 2015 \\ 
B1356-60 &   0.127 &   5.00 &  35.082 & $<$ 32.113 & $<$  -2.97  &  Prinz \& Becker 2015 \\ 
B1449-64 &   0.179 &   2.80 &  34.273 & $<$ 31.041 & $<$  -3.23  &  Prinz \& Becker 2015 \\ 
B1634-45 &   0.119 &   3.83 &  34.876 & $<$ 31.857 & $<$  -3.02  &  Prinz \& Becker 2015 \\ 
B1643-43 &   0.232 &   6.86 &  35.555 & $<$ 32.933 & $<$  -2.62  &  Prinz \& Becker 2015 \\ 
B1702-19 &   0.299 &   1.18 &  33.786 & $<$ 30.437 & $<$  -3.35  &  Prinz \& Becker 2015 \\ 
B1730-37 &   0.338 &   3.44 &  34.187 & $<$ 31.618 & $<$  -2.57  &  Prinz \& Becker 2015 \\ 
B1754-24 &   0.234 &   3.51 &  34.599 & $<$ 31.384 & $<$  -3.22  &  Prinz \& Becker 2015 \\ 
B1758-23 &   0.416 &   4.00 &  34.793 & $<$ 31.693 & $<$  -3.10  &  Prinz \& Becker 2015 \\ 
B1828-11 &   0.405 &   3.58 &  34.552 & $<$ 31.953 & $<$  -2.60  &  Prinz \& Becker 2015 \\ 
J0631+1036 &   0.288 &   6.54 &  35.240 & $<$ 31.887 & $<$  -3.35  &  Kennea et al., 2002, \\ 
J1055-6028 &   0.100 &  30.00 &  36.070 & $<$ 32.940 & $<$  -3.13  &  Prinz \& Becker 2015 \\ 
J1406-6121 &   0.213 &   9.11 &  35.349 & $<$ 32.557 & $<$  -2.79  &  Prinz \& Becker 2015 \\ 
J1412-6145 &   0.315 &   9.32 &  35.095 & $<$ 32.063 & $<$  -3.03  &  Prinz \& Becker 2015 \\ 
J1413-6141 &   0.286 &  11.00 &  35.751 & $<$ 32.230 & $<$  -3.52  &  Prinz \& Becker 2015 \\ 
J1514-5925 &   0.149 &   4.50 &  34.538 & $<$ 32.106 & $<$  -2.43  &  Prinz \& Becker 2015 \\ 
J1648-4611 &   0.165 &   5.71 &  35.319 & $<$ 31.262 & $<$  -4.06  &  Prinz \& Becker 2015 \\ 
J1723-3659 &   0.203 &   4.28 &  34.579 & $<$ 32.100 & $<$  -2.48  &  Prinz \& Becker 2015 \\ 
J1815-1738 &   0.198 &   9.01 &  35.595 & $<$ 33.029 & $<$  -2.57  &  Prinz \& Becker 2015 \\ 
J1828-1057 &   0.246 &   4.27 &  34.738 & $<$ 31.612 & $<$  -3.13  &  Prinz \& Becker 2015 \\ 
J1838-0549 &   0.235 &   4.73 &  35.005 & $<$ 31.727 & $<$  -3.28  &  Prinz \& Becker 2015 \\ 
J1850-0026 &   0.167 &  10.69 &  35.523 & $<$ 32.159 & $<$  -3.36  &  Prinz \& Becker 2015 \\ 
J1907+0918 &   0.226 &   7.68 &  35.508 & $<$ 31.064 & $<$  -4.44  &  Prinz \& Becker 2015 \\ 
J1908+0734 &   0.212 &   0.58 &  33.532 & $<$ 30.307 & $<$  -3.22  &  Prinz \& Becker 2015 \\ 
J1909+0749 &   0.237 &  10.84 &  35.653 & $<$ 33.131 & $<$  -2.52  &  this work \\ 

\hline  \multicolumn{7}{l}{Sample HB$_{\rm d}$ }  \\
B1509-58 &   0.151 &   4.40 &   37.242 &  34.938 &    -2.30&  Kargaltsev \& Pavlov 2008 \\ 
B1916+14 &   1.181 &   1.41 &   33.706 &  30.834 &    -2.87&  Zhu et al., 2009 \\ 
J0726-2612 &   3.442 &   3.01 &   32.453 &  32.812 & *    0.36&  Speagle et al., 2011 \\ 
J1119-6127 &   0.408 &   8.40 &   36.369 &  33.400 &    -2.97&  Safi-Harb \& Kumar 2008 \\ 
J1124-5916 &   0.135 &   5.00 &   37.077 &  33.422 &    -3.66&  Kargaltsev \& Pavlov 2008 \\ 
J1718-3718 &   3.379 &   5.08 &   33.217 &  32.571 & *   -0.65&  Zhu et al., 2011 \\ 
J1734-3333 &   1.169 &   7.40 &   34.750 &  32.425 &    -2.33&  Olausen et al., 2013 \\ 
J1819-1458 &   4.263 &   3.81 &   32.467 &  33.244 & *    0.78&  McLaughlin et al., 2007 \\ 
J1930+1852 &   0.137 &   7.00 &   37.063 &  34.123 &    -2.94&  Kargaltsev \& Pavlov 2008 \\ 

\hline  \multicolumn{7}{l}{Sample  HB$_{\rm ul}$ }  \\
B0154+61 &   2.352 &   1.61 &  32.758 & $<$ 30.285 & $<$  -2.47  &  Prinz \& Becker 2015 \\ 
B1610-50 &   0.232 &   7.24 &  36.196 & $<$ 33.242 & $<$  -2.95  &  Pivovaro et al., 1998 \\ 
J1524-5706 &   1.116 &  21.59 &  34.005 & $<$ 31.961 & $<$  -2.04  &  Prinz \& Becker 2015 \\ 
J1726-3530 &   1.110 &   9.97 &  34.547 & $<$ 32.113 & $<$  -2.43  &  Prinz \& Becker 2015 \\ 
J1841-0524 &   0.446 &   4.89 &  35.018 & $<$ 31.639 & $<$  -3.38  &  Prinz \& Becker 2015 \\ 
J1846-0257 &   4.477 &   4.69 &  31.850 & $<$ 29.603 & $<$  -2.25  &  Prinz \& Becker 2015 \\ 

\hline
\end{longtable*}

\section{Method of Analysis} \label{method}

The data set for  each pulsar contains
$(\Lx, \Lrot, d)$, where $\Lx$ and $\Lrot$ are measured in erg\;sec$^{-1}$,
and $d$ is in kpc.
Let $y = \log \Lx$, $a = \log \Lrot$, and
for the $i$-th pulsar in particular, denoting   $y_i$ and $a_i$.
We assume that there is an intrinsic model relation
between $\Lx$ and $\Lrot$, which is represented by
a linear formula,
\begin{equation} \eq{model1}
\ymod (a) = c_1 (a - 37 ) + c_2,
\end{equation}
where $c_1$ and $c_2$ are the constants.
The random variable we consider is the residuals defined by 
\begin{equation} \eq{random}
x = y - \ymod (a) .
\end{equation}

We use a Monte Carlo simulator which produces
a large number of  simulated $\Lx$ for a given pulsar and 
construct a probability density function $f(x)$.
The simulator takes into account possible effects 
which cause scatter in the $\Lx - \Lrot$ plot.
The detail of the simulation will be discussed in the next subsection.
If the model relation $\ymod$ is correctly guessed and
if the simulator reproduces the statistical characteristics of the 
{observation} data properly,
then the observation $x$ follows the probability distribution $f(x)$.
This is the hypothesis test to be performed.

\subsection{The Monte Carlo Simulator \label{simulator} }

The simulator works in the following way.

For a given pulsar with $a$,
the expected X-ray luminosity is given by 
$\Lx^{\rm model} = 10^{\ymod (a)}$ from (\ref{model1}).
The simulator produces residuals $x$ by using a random number generator
as described below.
Once $x$ is obtained, a simulated value of the X-ray luminosity is given by
$\log \Lx = x + y_{\rm model}$. 

The first step of the simulation is to include the geometrical effect.
If the radiation is isotropic from whole the star, 
one would simply observe the value $\Lx^{\rm model}$. 
However, if it is from a small hot area on the star,
one would observe $\Lx = \Lx^{\rm model} \cos \theta$,
where $\theta$ is the angle of the observer
to the normal of the emitting surface.
For a randomly oriented object, the probability density of observing
$\Lx$ is given by 
\begin{equation} \eq{eq4}
f(\Lx ) = 
\left\{
\begin{array}{lcc}
\displaystyle {1 \over 2  \Lx^{\rm model} } & \mbox{if } & 0 \leqq  \Lx \leqq  \Lx^{\rm model} \\ 
0                             & \mbox{if } & \Lx^{\rm model} < \Lx. 
\end{array}
\right.
\end{equation}
This simply means that $\Lx$ distributes uniformly
in between $\Lx^{\rm model}$ and zero when a hot spot
is observed by randomly distributed observers
(for derivation, see Appendix A).
Since the star rotates, the viewing angle $\theta$ oscillates through
one rotation, and therefore the value of $\theta$ is regarded as the
mean value.

\begin{figure*}
\begin{center}
\mbox{\raisebox{0mm}{ \includegraphics[width=7cm]{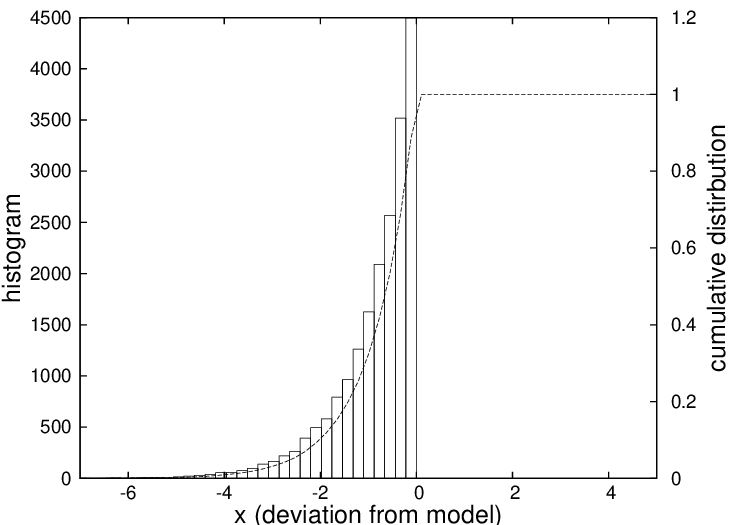} }}
\mbox{\raisebox{0mm}{ \includegraphics[width=7cm]{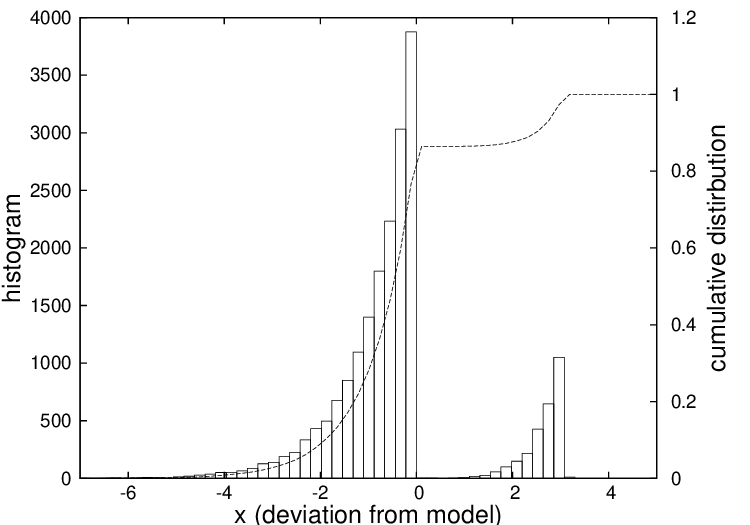} }}

\mbox{\raisebox{0mm}{ \includegraphics[width=7cm]{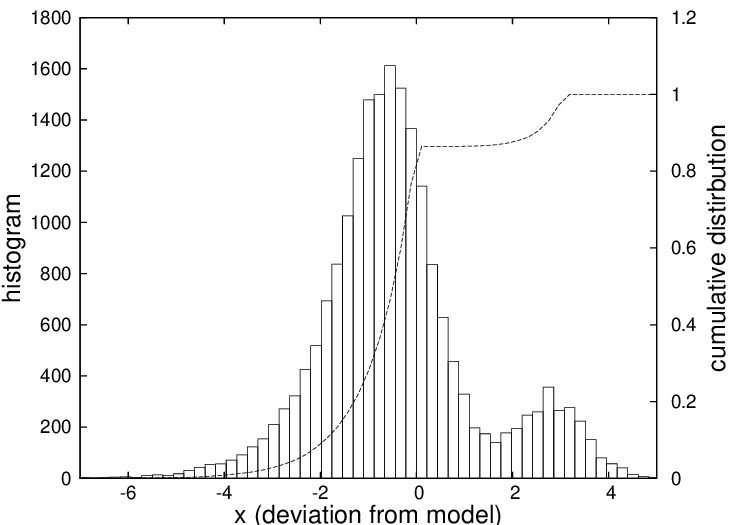} }}
\mbox{\raisebox{0mm}{ \includegraphics[width=7cm]{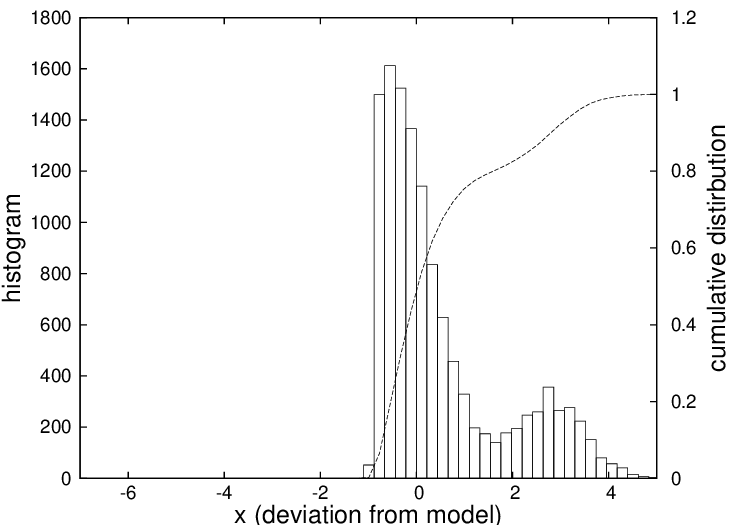} }}
\caption{ \label{rfig6}
The distributions of a simulated  $\Lx$ (histogram) and 
their cumulative distribution $g_i (x)$ (dotted curve) 
in each simulation step. 
The first step (upper left):including the effect of anisotropy,
the second step (upper right): adding the excess emission,
the third step (lower left): scattered by uncertain distance estimate and
interstellar absorption, and
the last step (lower right): 
the final distribution after the cut-off by the detection limit.
Note that the distributions are different object by object,
depending on $\Lrot$ and $d$, even if the simulation parameters are
the same.
}
\end{center}
\end{figure*}

The magnetospheric radiation would have a higher anisotropy along the
local magnetic field of the particle acceleration region.
A simple extension for geometrical  effect would be obtained
if we introduce an index $n$ and assume
$\Lx = \Lx^{\rm model} \cos^n \theta$.
In the simulator,
we model the effect of anisotropy in such a way that
$n=0$ for the isotropic radiation,
$n=1$ for the radiation from a small hot area on the star,
and $n >1$ for the magnetospheric directed radiation.
In this general case,
after transforming from $\Lx$ to $x$, we have
\begin{equation} \eq{view}
f( x ) = \left\{
\begin{array}{lcc}
(1 / n ) \exp \left( x / n \right)
& \mbox{if} & x \leqq 0 \\ 
0                             
& \mbox{if } & x > 0. 
\end{array}
\right.
\end{equation}
According to this probability, 
the simulator produces a number $x_{\mbox{\scriptsize I}}$,
which yields a X-ray luminosity affected by the geometrical effect
as $\log \Lx^{\rm psr} = {x_{\mbox{\scriptsize I}} + \ymod }$.
Thus dim pulsars are distributed below $y_{\rm model}$
(see the left top panel of Figure~\ref{rfig6}).

In the second step, we include the effect that
dissipation of the crustal magnetic field may add some amount of
X-ray luminosity.
However, we know little about the property of this kind of radiation.
We introduce two parameters:
(1) ${\cal P}_{\rm mag}$ is the probability that
such an excess emission appears, and 
(2) $\hat{L}_{\rm x}^{\rm mag}$ is the largest luminosity below which
the additional excess luminosity, $\Lx^{\rm mag}$, is uniformly distributed.
Again with the random number generator,
we find whether the excess radiation exists or not, and
if it exists, $\Lx^{\rm mag}$ is given. 
By adding the two components, we have 
a residual in the second step as 
\begin{equation}
x_{\mbox{\scriptsize II}} = \log (\Lx^{\rm psr} + \Lx^{\rm mag} ) - \ymod .
\end{equation}

In the third step, we consider uncertainties
in the estimated distance and interstellar absorption. 
The probability density function for this fluctuation is usually
assumed to follow a log-normal distribution.
He et al.(2013) examined the correlations among  $n_H$,
the dispersion measure $DM$ and the distances obtained by other methods.
From their Table~1,
we obtain a distribution of 
residuals from the linear regression $\log n_H = 0.3508 + \log DM$
 and
make the Kolmogorov-Smirnov (KS) test for it.
We find that the distribution fits very well with the log-normal 
distribution with the KS test statistics
$D=7.06 \times 10^{-2}$ and the significance level of 
${\cal P}_{\rm KS} =91.64$\%,
where we use {\sf ksone } given in ``Numerical Recipes''
(Press et al. 1992).
From the same table, we obtain the standard deviation for 
the difference between the dispersion measure distance and
the distances measured by parallax or HI absorption to be
$\sigma_{\rm dist} = 0.35$.

The scatter in $\Lx$ could also 
be caused by errors in
determining $n_H$ and consequently the unabsorbed flux.
This effect strongly depends on the statistical quality of
the spectral fitting for each pulsar.
The standard deviation of the errors in the unabsorbed flux 
in Kargaltsev \& Pavlov(2008) is found to be $0.14$ in log-scale.
This indicates that the scatter due to
 uncertainty in the absorption
is not as large as that of the distance squared,
whose standard deviation is $2 \sigma_{\rm dist} = 0.7$.

Regarding these points,
we assume a scatter $\delta$ according to a log-normal distribution
with standard deviation of $\sigma$
in the simulator.
The value of $\sigma$ is the model parameter,
but it is  constrained to be $\sigma \sim 0.7$.
We add $\delta$ for the third step: 
\begin{equation}
x_{\mbox{\scriptsize III}} = x_{\mbox{\scriptsize II}}+ \delta,
\end{equation}
by which a X-ray luminosity 
$ \log \Lx = x_{\rm III} + \ymod$ 
is obtained.

Finally, we take the observable flux limit into account.
In the simulator,
we use a parameter $F_{\rm lim}$ which defines the detection limit.
If the value of $\Lx$ obtained in the third step
is larger than $4 \pi d^2 F_{\rm limi}$, 
the final value $x_{\mbox{\scriptsize III}}$ is taken as an observation $x$, 
otherwise it is thrown out.

The above process is repeated $N$ times,
where $N$ is typically $2 \times 10^4$ for each pulsar to have stable results.
Thus the simulated values of $x$ 
yield the probability density function $f(x)$.

In summary, the model parameters are
$c_1$ and $c_2$ for the model relation,
$n$ for anisotropy,
${\cal P}_{mag}$ and $\hat{L}_{\rm x}^{\rm mag}$ for excess radiation by magnetic field decay,
$\sigma$ for scattering due to the distance estimate and the interstellar absorption,
and
$F_{\rm lim}$ for the detection limit of the instruments.
Typical distributions after each of the four steps is shown in Figure~\ref{rfig6}.

\subsection{Method of Statistical Test}

In this subsection, we describe
the statistical test we have used to see if a given sample follows 
a specific model.
For a given sample, we have the observations $\{ x_i, a_i, d_i \}$
where $i=1,2,...,N_{\rm psr}$, and
$N_{\rm psr}$ is the number of the pulsars.
For each pulsar,
the Monte Carlo simulator provides the probability density function 
$f_i (x)$.
It is notable that $f_i (x)$ is given for each pulsar,
which has the rotation power $a_i$ and the distance $d_i$ as known.
The simulator takes the values of $a_i$ and $d_i$ into account and applies
the assumed set of the model parameters
$\{ c_1, c_2, n, {\cal P}_{mag}, \hat{L}_{x}^{mag}, F_{lim} \}$.
To be exact, $f_i (x)$ may be written as
$f_i (x, a_i, d_i;n, {\cal P}_{\rm mag}, \hat{L}_{\rm x}^{\rm mag}, F_{\rm lim})$.

The random variable $x_i$ is now transformed into a new
variable $\xi_i$ by using the cumulative distribution,
\begin{equation}
g_i  (x) = \int_{- \infty}^x 
f_i (t, a_i, d_i;n, {\cal P}_{\rm mag}, \hat{L}_{x}^{\rm mag}, F_{\rm lim})
dt ,
\end{equation}
so that
the new random variable is given by $\xi_i = g_i (x_i)$.

The  distribution of $\xi_i$ becomes uniform
between 0 and 1 
if the observation follows the assumed mode.
Therefore,
we have made statistical tests for uniformity of $\{ \xi_i \}$.
We apply
the Kolmogorov-Smirnov (KS) test
and 
the $\chi^2-$test. 
KS test for $\{ \xi_i \}$ is straightforward by {\sf ksone} 
(Press et al. 1992),
while we 
provide $N_{\rm bin}$-histogram with respect to $\xi$ for
$\chi^2-$test.
After some trials, we find an appropriate bin size:
we used two sets of 7-bins (equally spaced in $\xi$) 
for $a > 35.5$ and for $a < 35.5$, respectively.
We have $N_{\rm bin}=14$
in total  (detail is given Appendix).

\subsection{Treatment for the Upper Limit Data}
In addition to the samples of the detected objects,
we also analyze the samples including the upper limit data.
If the upper limit is below the general trend, say  
$\Lx \sim (10^{-2} - 10^{-3}) \Lrot$,
they will  improve the statistics.
We use Sample SAB, i.e.,
Sample SAB$_{\rm d}$ and Sample SAB$_{\rm ul}$ are jointed.

We follow the general method to obtain a regression 
for the censored data as follows.
Since we treat the data below the detection limit,
we construct the  probability density function  that 
is obtained in the limit $F_{lim} \rightarrow 0$ by the simulator.
We denote it by $\hat{f}_i (x)$, 
and  its cumulative distribution by $\hat{g}_i (x)$.
If $x_i$ is the upper limit value, it is converted
to $\hat{\xi}_i = \hat{g}_i (x_i)$, below which
the actual value of $\xi_i$ should take.
Therefore, the observation $\xi_i$ is set to be a randomly chosen 
value in between zero and $\hat{\xi}_i$.
For all the objects in sample 
SAB$_{\rm ul}$, we obtain  a set of thus determined $\xi_i$.
For detected values in Sample SAB$_{\rm d}$, we simply have
$\xi_i = \hat{g}_i (x_i)$.
Note that this value is {\it not} $g_i (x_i)$ 
which is used previously for the detected values only.
Thus, we have a set $\{\xi_i \}$ for the joint Sample SAB
(Sample SAB$_{\rm d} +$ SAB$_{\rm ul}$).
The $\chi^2$ test and the KS test are applied to
$\{ \xi_i \}$ of Sample SAB in the same way used for Sample SAB$_{\rm d}$.

\section{Results} \label{result}

\begin{figure}
\begin{center}
\includegraphics[width=9cm]{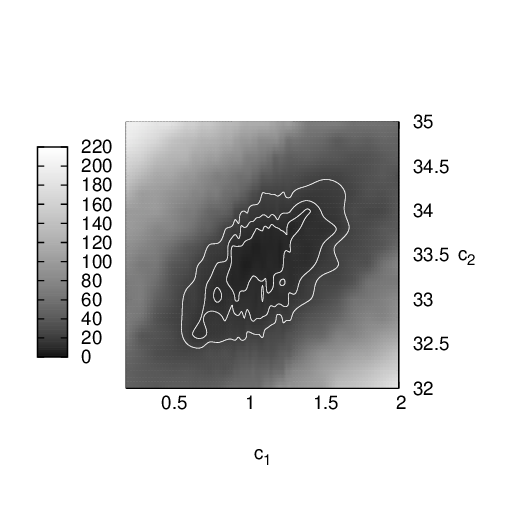}
\caption{ \label{rfig1}
The $\chi^2$ map for Sample SAB$_{\rm d}$.
The horizontal axis is the slope $c_1$  and the vertical axis is 
the normalization $c_2$.
The contours are drawn for the 1-$\sigma$ (68.3\%), 90\%, and 99\% 
confidence levels.}
\end{center}
\end{figure}

As the base line parameters of the simulator,
we take $n=3$,
${\cal P}_{\rm mag}=0$,
$\sigma = 0.7$,
and $\log F_{\rm lim} = -14.0$.
Actually this set is found to provide the best statistics
after some trials.
Using Sample SAB$_{\rm d}$,
we search for the most probable relation in the form of (\ref{model1}).
The result is shown in the $\chi^2$ map (Figure~\ref{rfig1}).
The contours are drawn for 1-$\sigma$ (68.3\%), 90\% and 99\% confidence levels.
We obtained the best fit model,
$c_1 = 1.03$ and $c_2=33.36$,
which is shown by the dashed line in Figure~\ref{LxLrot1}.
Since the $\chi^2$ map indicates a correlation between $c_1$ and
$c_2$ with a slope of 1.62, we find
\begin{equation} \eq{bestfit}
\log \Lx = c_1 (\log \Lrot - 35.38 ) + 31.69,
\end{equation}
with $c_1 = 1.03 \pm 0.27$.
For the best fit model, we obtain
$\chi^2/{\rm dof} = 9.036/13$ and
$D=0.10419$ with ${\cal P}_{\rm KS} =56.60$\%.

Since we use non-parametric test,
we also argue that 
the scatter around the regression line is reproduced well by our model. 
In a traditional way, the relation may be written as
$\Lx = 10^{-4.75} \Lrot^{1.03}$,
 or more roughly, 
since $c_1 \approx 1$, the X-ray efficiency is constant;
 $\log \eta_{\rm psr} = \log (\Lx / \Lrot) \approx -3.7$.

\begin{figure}
\begin{center}
\mbox{\raisebox{10mm}{ \includegraphics[width=6cm]{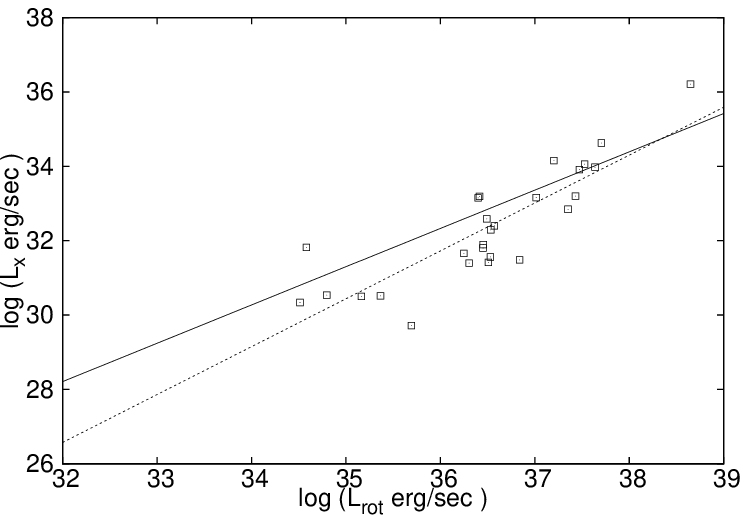} }}
\caption{ \label{rfig2}
The $\Lx - \Lrot$ plot for Sample S$_{\rm d}$.
The solid line indicates the best fit relation, while
the dotted line does the  apparent regression line.}
\end{center}
\end{figure}

\begin{figure}
\begin{center}
\mbox{\raisebox{10mm}{ \includegraphics[width=6cm]{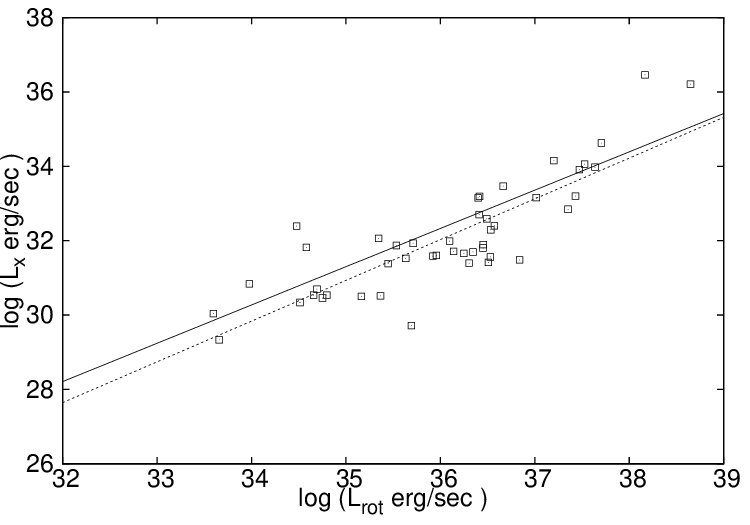} }}
\caption{ \label{rfig3}
The same as Figure~\ref{rfig2} but for the joint sample of
S$_{\rm d}$ and A$_{\rm d}$. }
\end{center}
\end{figure}

Let us consider dependence of the result on samples.
We test the best fit model $(c_1 , c_2) =(1.03, 33.36)$ on
the two samples,
Sample S$_{\rm d}$ and the joint sample S$_{\rm d} + $A$_{\rm d}$. 
We find that the best fit model fits very well to the both samples.
For Sample S$_{\rm d}$,
we have
$\chi^2/{\rm dof} = 6.091/13$ and
$D=0.09569$ with ${\cal P}_{\rm KS} =96.57$\% ,
and 
for Sample S$_{\rm d}+$A$_{\rm d}$,
$\chi^2$/dof = 5.424/13, and 
$D= 0.099$ with ${\cal P}_{\rm KS} =74.7$\%.
The best fit model is drawn by the solid line 
in Figure~\ref{rfig2} for Sample S$_{\rm d}$, and 
in Figure~\ref{rfig3} for Sample S$_{\rm d}+$A$_{\rm d}$.
In these figures, the regression lines obtained by the usual method
are shown by the dotted lines.
It is remarkable that the apparent regression lines 
dotted lines are 
different sample by sample, 
while the intrinsic relation is unchanged. 
This means that 
the apparent regression lines are influenced by selection effect,
which is absorbed by the simulator, so that
the both sample are best fitted by the single model.

There are two factors that affect the apparent regression line. 
Pulsars with large $\Frot$ 
have high probability to be observed even if they 
have low luminosities.
Such  pulsars are present in 
$10^{35} \lesssim \Lx \lesssim 10^{37}$ in 
Figure~\ref{rfig2}
distributing below the best fit $\Lx - \Lrot$ relation.
This effect makes the regression line steeper as seen in 
Sample S$_{\rm d}$. 
The same would occur if still deeper observations were made
in future. 
On the other hand,
if pulsars with small $\Frot$  are included, say by adding Sample B or C,
then the instrumental flux limit becomes important; namely
dim pulsars such as observed in Sample S 
are difficult to be observed.
Therefore, by adding data for pulsars with small  $\Frot$
(small $\Lrot$ and large distances), pulsars with large $\Lx$ are
selectively included in  the sample. 
This makes the  apparent regression line flatter.
This can be seen in Figure~\ref{rfig3} 
(Sample S$_{\rm d}+$A$_{\rm d}$ ), i.e.,
more data appear above the regression line in the low $\Lrot$ regime
($10^{33} \lesssim \Lrot \lesssim 10^{35}$). 
Therefore, the slope of the  apparent regression line
becomes  flatter in 
Sample S$_{\rm d} +$ A$_{\rm d}$.
It is therefore notable that
 the slope of a apparent regression line is affected
by properties of samples, and the simulation of the probability density $f_i(x)$ including
individuality must be stressed.

Next we examine the dependence of the simulation parameters.
The best fit model relation (\ref{bestfit}) 
is fixed, and the simulation parameters,
$n$, ${\cal P}_{\rm mag}$, $\sigma$, and $F_{\rm lim}$,
are changed separately to see how the test statistics changes.
The result is summarized in Table~\ref{para}.

Regarding the anisotropy parameter $n$,
we find that on average the X-ray radiation comes
neither from the whole neutron star nor from the hot region on the
surface, and that
anisotropy with $n \sim 2 - 3$ is preferred.
The model of hot spot ($n=1$) does not have a good fit to the data,
though  the test statics is marginal 
(${\cal P}_{\rm KS} = 14.48$\%). 
The result suggests that
the X-ray radiation is beamed, rather than from a hot spot,  for 
at least for some important fraction of the pulsars.

\begin{table*}
\caption{ \label{para}
The result of $\chi^2-$test and KS test for different simulation parameters. 
The degree of freedom of $\chi^2$ 
is 13.
}
\begin{center}
\begin{minipage}{0.3\hsize}
\begin{tabular}{crr} \hline \hline
$n$ & $\chi^2$ & ${\cal P}_{\rm KS}$ \\ 
    &          & ( in \% ) \\\hline
0 &          30.418 & 0.00 \\
1 &          10.236 & 14.48 \\
2 &           9.036 & 56.60 \\
3 &          12.545 & 42.19 \\
4 &          12.873 & 34.04 \\
6 &          15.673 & 31.46 \\
\hline 
\end{tabular}
\end{minipage}
\begin{minipage}{0.3\hsize}
\begin{tabular}{crr} \hline \hline
$\sigma$ & $\chi^2$ & ${\cal P}_{\rm KS}$\\ 
    &          & ( in \% ) \\\hline
0.5 &          19.964 & 7.70 \\
0.6 &          16.600 & 25.97 \\
0.7 &           9.036 & 56.60 \\
0.9 &          10.164 & 19.30 \\
1.0 &          12.545 & 10.83 \\
1.3 &          16.291 & 2.48 \\
\hline
\end{tabular}
\end{minipage}
\begin{minipage}{0.3\hsize}
\begin{tabular}{crr} \hline \hline
$\log F_{\rm lim}$ & $\chi^2$ & ${\cal P}_{\rm KS}$\\
    &          & ( in \% ) \\\hline
-13.0 &          97.600 & 0.00 \\
-13.5 &          45.218 & 0.00 \\
-14.0 &           9.036 & 56.60 \\
-14.5 &          19.491 & 6.02 \\
-15.0 &          24.109 & 0.30 \\
\hline
      &                 &       \\
\end{tabular}
\end{minipage}
\end{center}

\end{table*}

The scatter due to the
uncertainty in distance and interstellar 
absorption is consistent with the estimate $\sigma = 0.7$,
which is suggested by He et al.(2013)
(see \S~\ref{simulator}).
The acceptable value would be at most $\sigma \sim 0.9$
(see Table~3).

The detection limit is very much sensitive to the fitting.
The acceptable value is very narrow.
If $\log F_{\rm lim}  \gtrsim  -13.5$, no acceptable model is found.
We argue that regarding the detection limit of the instruments is 
very much important in a statistical study of $\Lx$.

\begin{figure}
\begin{center}
\mbox{\raisebox{-9mm}{ \includegraphics[width=8cm]{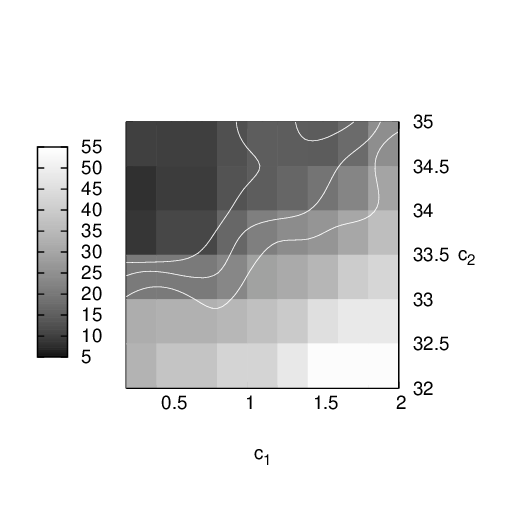} }}
\caption{ \label{rfig5}
The same as Figure~\ref{rfig1} but for Sample HB. }
\end{center}
\end{figure}

We examine whether the high-magnetic field pulsars, Sample HB, follows
the best fit model.
The result is summarized in Table~\ref{HB}.
If we assume ${\cal P}_{\rm mag} = 0$,
the best fit parameters  obtained for Sample SAB 
gives the statistics as
${\cal P}_{\rm KS} = 0.09$\% for Sample HB$_{\rm d}$ and
${\cal P}_{\rm KS} = 4.65$\% for the joint sample HB $+$ SAB.
Therefore, 
non existence of the excess emission is rejected for the both samples. 
In spite of the fact that sample
HB is a subset of the rotation powered radio pulsars, 
they do not obey the best fit model  for Sample SAB$_{\rm d}$.
As far as Sample HB$_{\rm d}$ is concerned,
the $\chi^2$ map in Figure~\ref{rfig5} indicates $c_1 \sim 0$,
i.e., no correlation with respect to $\Lrot$.
It may be argued from Table~\ref{HB} that
${\cal P}_{\rm mag} \gtrsim 0.1$ for the high-magnetic field pulsars,
and
${\cal P}_{\rm mag} \sim 0.05$ for the joint sample. 
This is just a reflection of the fact that
out of 9 pulsars in Sample HB$_{\rm d}$, 3 pulsars have large efficiency
$\Lx /\Lrot > 0.1$. 

\begin{table}
\caption{ \label{HB}
$\chi^2$ and KS significance level for the sample including
the high-magnetic field pulsars.
}
\begin{center}
\begin{minipage}{0.3\hsize}
\begin{tabular}{crr} 
\multicolumn{3}{c}{Sample HB} \\
\hline \hline
${\cal P}_{mag}$ & $\chi^2$ & ${\cal P}_{\rm KS}$ \\ 
    &          & ( in \% ) \\\hline
0.0 &          23.90 &    0.09 \\
0.05&          18.30 &    7.80 \\
0.1 &          18.30 &   32.75 \\
0.2 &          14.80 &   46.56 \\
0.3 &          12.00 &   50.43 \\
0.5 &           8.50 &   57.93 \\
1.0 &           8.50 &   59.83 \\
\hline 
    &                 &         \\
\end{tabular}
\end{minipage}
\hspace{20mm}
\begin{minipage}{0.3\hsize}
\begin{tabular}{lrr} 
\multicolumn{3}{c}{Sample HB $+$ SAB} \\
\hline \hline
${\cal P}_{mag}$ & $\chi^2$ & ${\cal P}_{\rm KS}$ \\ 
    &          & ( in \% ) \\\hline
0.0  &          13.85 &    4.65 \\
0.001 &          7.63 &   15.88 \\
0.01 &           8.67  &   47.77 \\
0.02 &           7.63 &   76.99 \\
0.05 &           6.71 &   92.54 \\
0.1  &           9.50 &   72.80 \\
0.2 &           18.24 &   17.54 \\
0.4 &           24.54 &    0.21 \\
\hline
\end{tabular}
\end{minipage}
\end{center}
\end{table}

\begin{figure}
\begin{center}
\mbox{\raisebox{-9mm}{ \includegraphics[width=8cm]{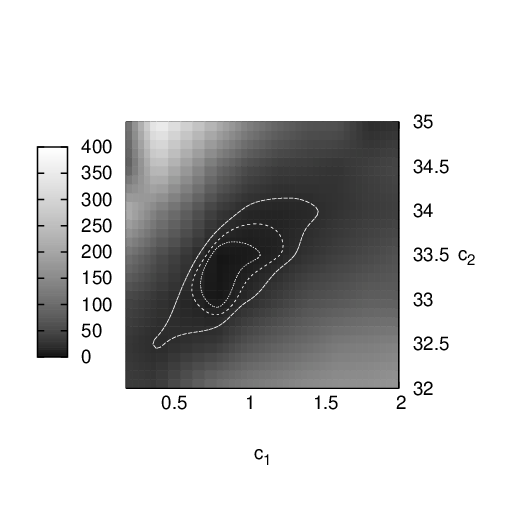} }}
\caption{ \label{rfig4}
The same as Figure~\ref{rfig1} but for the joint sample of
S, A and B including the upper limit data. }
\end{center}
\end{figure}

If we add 
the upper limit data, Sample SAB$_{\rm ul}$,
to Sample SAB$_{\rm d}$,
the number of the pulsar  is 99 in total (see Table~1).
The $\chi^2$ map for this joint sample is shown in Figure~\ref{rfig4}.
The best fit model  gives the test statistics,
$\chi^2/{\rm dof} = 14.091/13$ and
$D=0.09681$ with ${\cal P}_{\rm KS} =31.15$\%.
From the $\chi^2$ map, the most probable $c_1$ get slightly smaller
and $c_2$ is slightly higher.
The result of the sample including the upper limit 
yields $c_1 \approx 1.0$ and $c_2 \approx 33.5$,
of which the test statistics is
$\chi^2/{\rm dof} = 12.75/13$ and
$D=0.05392$ with ${\cal P}_{\rm KS} = 93.57$\%,
in very good agreement with the result of Sample SAB$_{\rm d}$.

\section{Discussion} \label{discussion}

To understand the distribution in the $\Lx - \Lrot$ plot,
we take the following effects into account:
(1) anisotropic radiation with randomly oriented viewing angles,
(2) uncertainty in the distance estimate, and
(3) detection limit mainly determined by the instruments.
These effects obscure a possible intrinsic relation between
$\Lx$ and $\Lrot$.
Regression lines which are obtained by the usual way 
are in general found to be different from the intrinsic $\Lx - \Lrot$ relation
due to selection effect. 
Regarding the above effects with the Monte Carlo simulator,
we have obtained the best fit model relation 
$\Lx = 10^{-4.75} \Lrot^{1.03}$. 
The scatter about the model relation is reproduced well
by  the Monte Carlo simulator; the $\chi^2$ and KS test give very
good statistics for the distribution.

There are three parameters, $n$, $\sigma$, and $F_{\rm lim}$
to reproduce the distribution.
However,   
$\sigma \sim 0.7$ and $F_{\rm lim} \sim 10^{-14}$~erg~cm$^{-2}$s$^{-1}$,
so that only $n$  can be
We find the most probable value $n \sim 2$
(anisotropy with $F_x \propto \cos^2 \theta $.) 

The above analysis is established for the samples in which the high-magnetic
field pulsars ($B_d > 10^{13}$~G) are excluded.
For Sample HB (high-magnetic field pulsar only), 
$\Lx$ seems not to correlate with $\Lrot$ at all.
For the joint sample, Sample HB$+$SAB,
statistically acceptable models are obtained only if
we introduce non-zero ${\cal P}_{mag}$; namely
there must be a finite  probability of the excess X-ray radiation.
The present Sample HB$+$SAB gives ${\cal P}_{mag} = 0.005$.
The most likely source of the radiation would be magnetic field decay.
In spite of the fact that the high-magnetic field pulsars emit radio pulses in the same
way as the ordinary radio pulsars, they form a distinctive sub-class
in the sense that they do not follow the model $\Lx - \Lrot$ relation that is established for
the ordinary pulsars.
Three high-magnetic field pulsars, which show the excess emission,
J0726-2612, J1718-3718, and J1819-1458,
possess  very small values of $\Frot$, which are respectively
$\log \Frot = -12.533$,
$-11.793$, and $-12.240$.
This means that Sample HB suffers strong selection effect.
We cannot conclude the
true probability ${\cal P}_{mag}$ of magnetic heating at present.
It will be obtained
if we could have
a complete sample within a given volume.

\begin{figure}
\begin{center}
\mbox{\raisebox{-9mm}{ \includegraphics[width=6cm]{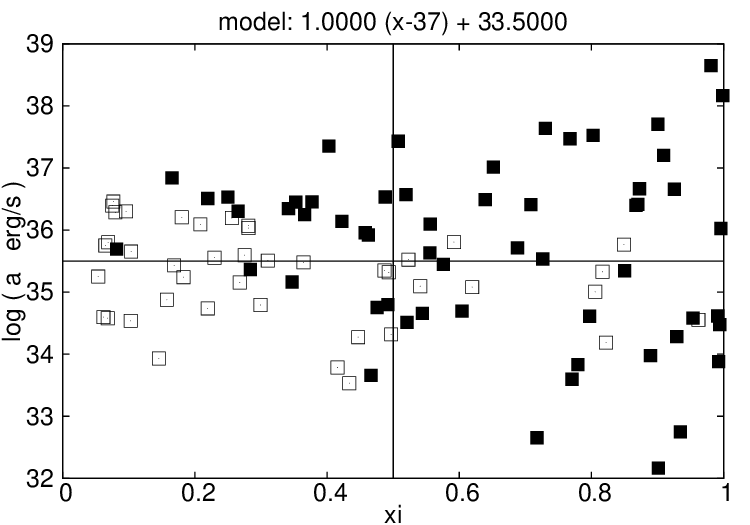} }}
\caption{ \label{rfig44} The scatter plot $(\xi, a)$ for Sample SAB with
the best fit model.
The data of finite detection are indicated by the filled squares, while
the upper limit data are indicated by the open square.} 
\end{center}
\end{figure}

Let us next consider the reason why some pulsars in Sample SAB
show high X-ray efficiency. 
To this end, we list the pulsars whose $\xi$ is lager than $0.9$:
we have 10 pulsars in Table~5.
Because the value of $10^{13}$~G used to define Sample HB
is rather ad hoc,
there may exist a pulsar showing an magnetic heating in 
the high-$\xi$ pulsars.
Another possibility is that
there is unknown physics which makes $\xi$ large.
In addition to Table~\ref{apu1},
we also provide a $(\xi_i, a_i)$ plot for Sample SAB 
in Figure~\ref{rfig44},  which may be helpful
to understand the distribution with respect to $\xi$.

Figure~\ref{rfig44} shows that
high-$\xi$ pulsars distribute for all range of $\Lrot$.
The only exception is
two very energetic pulsars,
PSR~B0540-69 and PSR~B0531+21 (Crab),
 with $\Lrot > 10^{38}$erg~s$^{-1}$
(indicated by "VE" in the Table~\ref{apu1}).
This may imply that the linearity of the 
$\Lx- \Lrot$ relation might not be hold 
for very energetic pulsars.
But, the number of samples is too small to
make the conclusion.

Of the remaining 8 pulsars, four pulsars have relatively large
$\log \Frot \gtrsim -9$, while
the remaining four have small 
$\log \Frot \lesssim -10$
(indicated by "H" and "L" in Table~5, respectively).
In the large $\Frot$ subset,
PSR~J1617-5055 and PSR~J1400-6325 are observed
in the hard X-ray bands ($> 20$~keV)
and classified as the soft gamma-ray pulsars (Kuiper \& Hermsen, 2015).
Photon indices are $\sim 1$ indicating their luminosity dominates in
hard X and soft gamma-ray bands.
However, neither are  
detected with {\it Fermi} LAT so that there must
be a cut off in somewhere in MeV bands (Kuiper \& Hermsen, 2015).
The soft gamma-ray pulsars become high $\xi$ pulsars because of their characteristic
spectral energy distribution. 
On the other hand,
PSR~B0656+14 and PSR~J1741-2054 are bright in thermal emission.
These two with PSR~B1055-52 are known as the neutron stars with high surface temperature 
in comparison with the standard cooling curve 
(Yakovlev et al. 2011, Karpova et al.,2014).
For these pulsars, the spectrum can be fitted by one or two back-body with
a less-luminous hard component fitted by
the power low with indices typically of $\sim 2$.

Among the four small $\Frot$ pulsars,
PSR~B1055-52 is a bright source and studied well
(De Luca et al. 2005, Posselt et al. 2015).
Its X-ray spectrum is well fitted by two black body models plus a power low model.
The dominant component is the black body with the temperature of 68~eV.
PSR~J0855-4644 shows a power law spectrum with photon index of 1.24, indicating 
this pulsar would belong to the soft gamma-ray type.
For PSR~B1822-14 and PSR~J1301-6310, one needs better observational data
to have finer spectral properties.

Although the number of sample is small,
we can recognize two types of pulsars showing large X-ray efficiency.
One is the soft gamma-ray type for which
the rotational luminosity is dominated in the soft gamma-ray and hard X-ray
bands.
The other is the thermally bright type,
which shows a high surface temperature as compared with the 
standard cooling curve.
The luminosity is thought to originate
for which the luminosity originates from the neutron star
with additional heating or suppressed cooling 
(Gusakov et al. 2004, Page et al. 2004).
However, noticing that 
the high-magnetic field pulsars do not follow the $\Lx - \Lrot$ relation due
to extra heating by the magnetic field and 
there are magnetars with small $B_d$,
we suggest that
the thermally bright type pulsars
with high $\xi$ are candidate objects that own dissipative magnetic
filed like magnetars.
The objects plotted as magnetars in Figure~\ref{LxLrot1} 
are in their active states.
After the active phase or outbursts,
majority of magnetars may reside in the distribution of the ordinary radio pulsars and
may show high $\xi$.

\begin{table*}
\begin{center}
\caption{ \label{apu1} Large efficiency  pulsars $\xi > 0.9$. }
\begin{tabular}{llllllllll}  \hline
pulsar name & $\xi $ & $\log \Lrot$ & $\log \Frot$ & Fermi & Hard X  & Type   & \multicolumn{2}{c}{Spectral} & PWN \\
            &         &              &              & LAT   &        &    & \multicolumn{2}{c}{Properties} & $\eta_{\rm pwn}$    \\ 
            &         &              &              &       &        &     & BB & PL          &     \\ \hline
  B0540-69 & 0.998 & 38.17 & -9.37 &    no &   Yes &    VE &          - &   PL(2.05) & -0.89 \\ 
J0855-4644 & 0.994 & 36.02 & -10.04 &    no &     - & L,nth &          - &   PL(1.24) & -2.01 \\ 
  B1055-52 & 0.992 & 34.48 & -9.97 &   Yes &     - &  L,th & 2BB(68~eV) &          - & -5.28 \\ 
  B0531+21 & 0.986 & 38.65 & -6.03 &   Yes &   Yes &    VE &          - &   PL(1.63) & -1.64 \\ 
  B0656+14 & 0.969 & 34.58 & -8.39 &   Yes &     - &  H,th & 2BB(56~eV) &          - & -5.68 \\ 
J1301-6310 & 0.956 & 33.88 & -10.82 &    no &     - &   L,? &          ? &          ? & - \\ 
  B1822-14 & 0.931 & 34.61 & -10.93 &    no &     - & L,th? & BB(200~eV) &          - & - \\ 
J1617-5055 & 0.922 & 37.20 & -8.49 &   Yes &   Yes & H,nth &          - &   PL(1.15) & -3.41 \\ 
J1400-6325 & 0.918 & 37.70 & -8.06 &    no &   Yes & H,nth &          - &   PL(1.22) & -2.76 \\ 
J1741-2054 & 0.901 & 33.98 & -9.05 &   Yes &     - &  H,th &  BB(60~eV) &          - & -4.58 \\ 

\hline
\end{tabular}
\end{center}
\end{table*}

\begin{figure}
\begin{center}
\includegraphics[width=8cm]{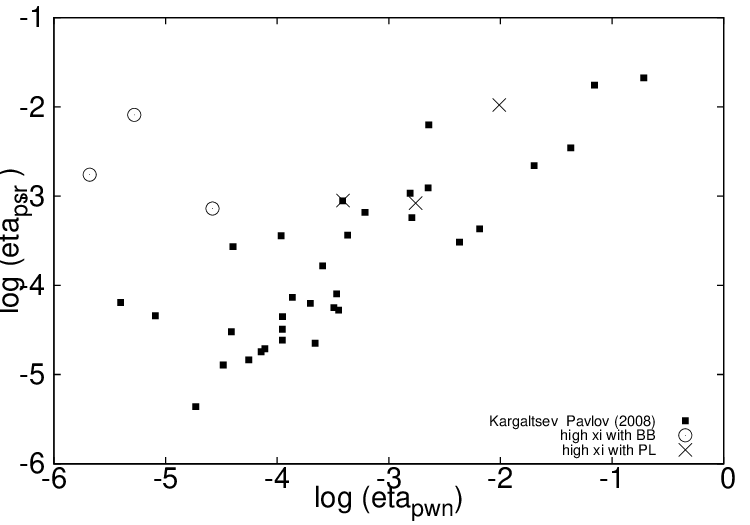}
\caption{ \label{figpu1}
Correlation between the pulsar and pulsar wind nebula efficiency.
The date of Kargaltsev \& Pavlov(2008) are indicated by the filled squares.
The high $\xi$ pulsars of the thermally bright type (the open circles)
and the soft gamma-ray type (the crosses) are also plotted.}
\end{center}
\end{figure}

It has been suggested (Kargaltsev et al. 2012)
and quantitatively confirmed by our analysis that
the probability density distribution $f(x)$ 
extends to smaller values of $x$; namely,
some pulsars appear very dim as compared with the model $\Lx - \Lrot$ relation.
In our model, we simulate the distribution by the exponential form 
$f(x)=(1/n)\exp (x/n)$ as shown
in the top left panel of Figure~\ref{rfig6}.
The model is drawn as the geometrical  effect.
However, the reason of the extended distribution
can be different. 
What we show is  that if the distribution is assumed in this form,
then the observed scatter is reproduced.
As is pointed out by Kargaltsev \& Pavlov(2008), 
similar things are found for the luminosity of PWN, i.e.,
some pulsar shows very small efficiency of the nebula emission.
Vink et al. (2001) argue that
the X-ray efficiency of PWN and that of the pulsar show similar behavior
if they are plotted against the spin-down age.
Taken from Table~2 of Kargaltsev \& Pavlov(2008), 
the efficiencies of the pulsar and PWN,
$\eta_{\rm psr} = \Lx /\Lrot$ and 
$\eta_{\rm pwn} = L_{\rm pwn} /\Lrot$, 
are plotted
in Figure~{\ref{figpu1}}.
Since the brightness of PWN is less dependent on
 the viewing angle,
the wide distribution in $\eta_{\rm pwn}$ must not be caused by
the viewing angle, but must be due to some unknown physics.
It is noticeable that $\eta_{\rm psr}$ and $\eta_{\rm pwn}$ is
positively correlated.
This indicates that 
there is obviously at least one parameter other than $\Lrot$, in other words, 
some unknown physics that governs the luminosity
of both the pulsar and PWN.
A possible link between the magnetospheric emission and the pulsar wind
is pair multiplicity.
If pairs are created efficiently, then the synchrotron emission from the
magnetosphere in X-ray would be enhanced, 
and at the same time, the kinetic part of the energy 
carried by the wind 
would increase and causes a brighter PWN.
This view is consistent with the fact that the 
$L_\gamma - \Lrot$ correlation is tighter,
i.e., the gamma-ray comes not from the secondary pairs but
from the primary particles.

The high-$\xi$ pulsars
are also plotted in Figure~\ref{figpu1}.
The soft-gamma type with high $\xi$  
follows the general trend
(indicated by the crosses in Figure~\ref{figpu1}).
The thermally bright  pulsars show 
small $\eta_{\rm pwn}$ and
large $\eta_{\rm psr}$, i.e., they do not follow the general trend
(indicated by the open circles in Figure~\ref{figpu1}).
For these, although $\eta_{\rm psr}$ is large, the luminosity originates 
from the heat of the neutron star, and the efficiency of the magnetospheric emission is
small so that the correlation holds even for these pulsars.

Further statistical analysis with much better quality of data,
separated into thermal, magnetospheric and PWN components, will give us
finer discrimination of individual origins of emission, and
a hint to find the unknown physics controlling the X-ray efficiencies.

We exclude MSPs from the samples. 
In the next step, we examine whether $\Lx - \Lrot$ plot of MSPs differs from that of the
ordinary pulsars. 
The weak dipole field or small curvature radius of MSP
may cause different dependence of pair creation rate on $\Lrot$ or other parameters.
`Buried' magnetic field by accreting matter may cause  an additional heating.
We may have a hit of these effect in the comparison.

\acknowledgments
The present work was supported in part 
by a Grant-in-Aid for Scientific Research (S.S. 25400221,
AB 15K05107, TE 15H00845) from the MEXT.
We thank H. Ohno for his helpful discussions  and
R. Shannon for carefully reading the draft and his comments.

\appendix
\section{Modeling of Geometrical Effects}

A simple model for the geometrical effect is
obtained 
if we consider the case in which 
a small hot spot on the stellar surface is observed.
Let the position vector of the spot, the observer's direction and
the angle between the two be respectively $\vec{R}$, 
$\vec{i}$, 
and $\theta$.
The observed flux may be given by $\Fx = F_0 \cos \theta$,
where $F_0$ is the observed flux when $\theta =0$.
Here we ignore the general relativistic effect.
If we take $\Fx$ as an random variable, then the
the probability distribution function $f(\Fx )$ is defined such that
the chance probability of observing the flux in between $\Fx$ and 
$\Fx + d \Fx$ is
\begin{equation}
{\rm Pr.} = f(\Fx) d \Fx = f(\Fx ) F_0 d(\cos \theta).
\end{equation}
On the other hand, the probability for the spot to locate in between $\cos \theta$
and $\cos \theta + d (\cos \theta )$ is given by
\begin{equation}
{\rm Pr. } = {1 \over 4 \pi } \left[ d(\cos \theta ) \int_0^{2 \pi} d \phi 
\right] = { d (\cos \theta ) \over 2},
\end{equation}
provided that the spot is randomly distributed on the surface.
Comparing the two expression, we have,
for $0 \leqq \Fx \leqq F_0 $,
\begin{equation}
f(\Fx ) = {1 \over 2F_0} 
\end{equation}
else $f(\Fx ) =0$, 
i.e., $\Fx$ distribute uniformly below $F_0$.
Note that $\int f(\Fx ) d \Fx = 1/2$ because spots on the backside of the
start would not be observed.
The generalized  expression is $\Fx = F_x \cos^n \theta$
with the anisotropy parameter $n$,
where $n$ larger than unity indicates the radiation is beamed.
In the same way, we have,
for $0 \leqq \Fx \leqq F_0 $,
\begin{equation} \eq{A14}
f(\Fx ) = {1 \over 2nF_0}  \left( \Fx \over F_0 
\right)^{ {1 \over n} - 1}.
\end{equation}
In general, the observer's direction has a finite angle to the emitting 
direction $\vec{R}$ so that the observed flux tends to smaller than $F_0$.
In the Monte Carlo simulation $F_0$ is 
replaced by  $\Lx^{\rm model}$, below which.
$\Lx$ is distributed according to the probability (\ref{A14}).
The distribution (\ref{A14}) can be seen in the $\Lx - \Lrot$ plot 
as  some dim pulsars are found
below an expected correlation.

If the viewing angle to the rotation axis were given
for each pulsar,
a correction might be possible.
However, we do not have convincing values of the viewing angles so that
such a correction is difficult to made.

In the simulation, the random variable $x=\log \Fx - \log F_0$ is used.
The probability distribution function with respect to $x$ and its cumulative
distribution become, respectively,
\begin{eqnarray}
f(x) &=& { \ln 10 \over n } 10^{x/n} \\
G(x) &=& \int_{- \infty}^x f(x^\prime ) dx^\prime
= 10^{x /n}.
\end{eqnarray}
The random values which follow $f(x)$ are produced  by
random numbers $G$, which is distributed
 uniformly between one and unity, with
\begin{equation}
x = n \log G .
\end{equation}

The viewing angle actually changes due to rotation according to
\begin{equation}
\cos \theta (t) =
\cos \theta_0 \cos \zeta - \sin \theta_0 \sin \zeta \sin \Omega t ,
\end{equation}
where $\vec{\Omega}$ is the angular velocity of the star,
$\cos \zeta = \vec{i} \cdot \vec{\Omega}$ indicates the observers direction
and $\Omega = |\vec{\Omega}|$.
Since we treat the phase averaged flux, the mean value of $\cos^n \theta (t)$
should be used to evaluate the effective value of $\theta$ 
in the Monte Carlo simulator.
However, we simply  assume the randomly distributed observer and use
the distribution (\ref{A14}).

\section{Spectral Analysis of PSR~J1909+0749}

PSR~J1909+0749 was observed serendipitously 
by {\it Chandra} on 2008 (ObsID 9614), February 28 
using Advanced CCD Imaging Spectrometer (ACIS).
The data reduction was done 
with the {\it Chandra} Interactive Analysis Observations (CIAO) software (version 4.7).
The radio pulsar was imaged on the ACIS-S2 chip, $13^\prime$ off-axis.
No source was found on the chip by the CIAO {\tt wavdetect} script.
Therefore, we derived an upper limit on the count rate.
To generate a new level-2 event file for the observation data, 
we made use of {\tt Chandra\_repro} preprocessing script.
Next, we performed time filtering using {\tt dmgti} and {\tt deflare} scripts,
and then the exposure-corrected image was created by {\tt fluximage} scripts.
We calculated the count rate by exposure-corrected image 
with {\tt srcflux} scripts. 
The count rate was converted into the unabsorbed flux 
by using {\tt PIMMS}, for which $N_{H}$ value is obtained by 
$
N_{H}(10^{20} {\rm cm^{-2}})=0.30^{+0.13}_{-0.09}{\rm DM}({\rm pc\ cm^{-3}})
$
(He et al. 2013) and  we used a power law model with a photon index of 1.5.

\section{Improvement of $\chi^2-$test }

\begin{figure*}
\begin{center}
\mbox{\raisebox{-18mm}{ \includegraphics[width=6cm]{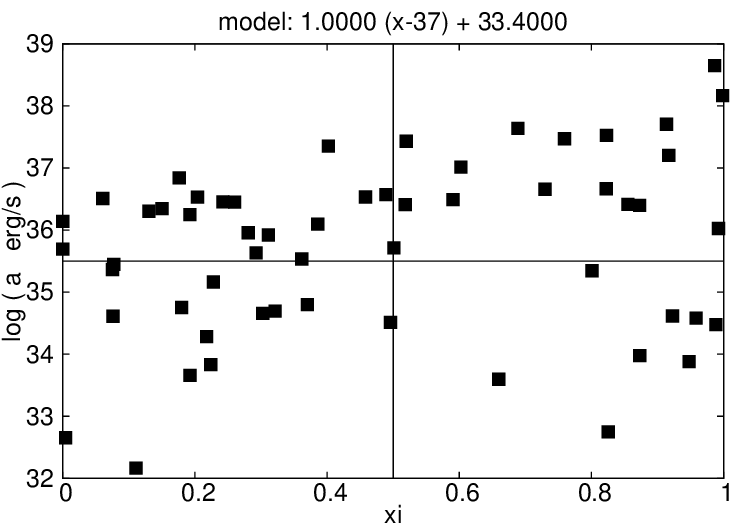} }}
\begin{minipage}{6cm}
\includegraphics[width=58mm]{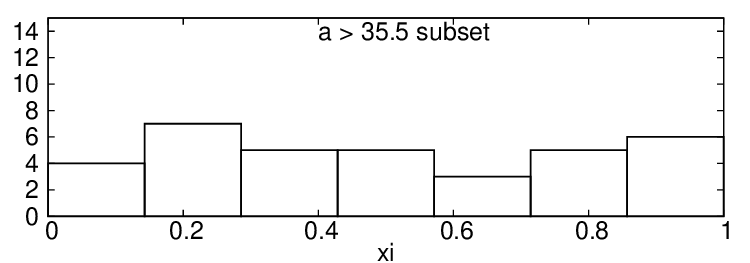} \par
\includegraphics[width=58mm]{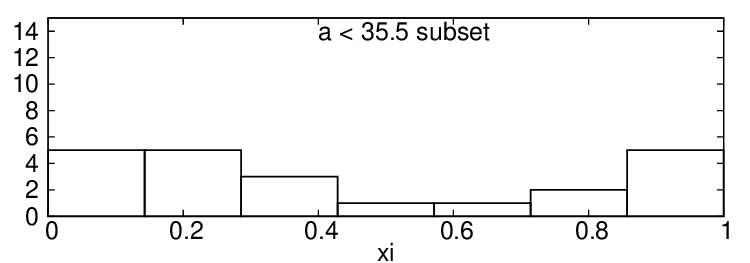}
\end{minipage}

\mbox{\raisebox{-18mm}{ \includegraphics[width=6cm]{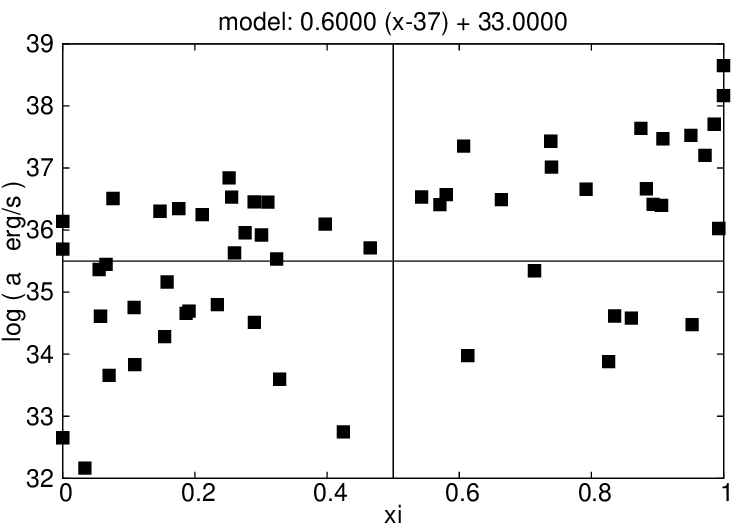} }}
\begin{minipage}{6cm}
\includegraphics[width=58mm]{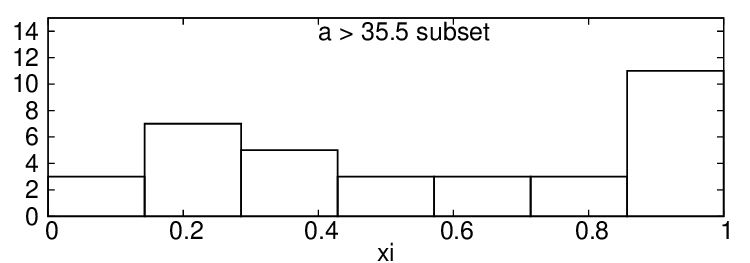} \par
\includegraphics[width=58mm]{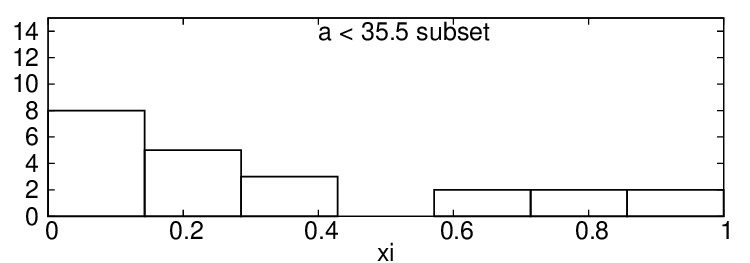}
\end{minipage}

\mbox{\raisebox{-18mm}{ \includegraphics[width=6cm]{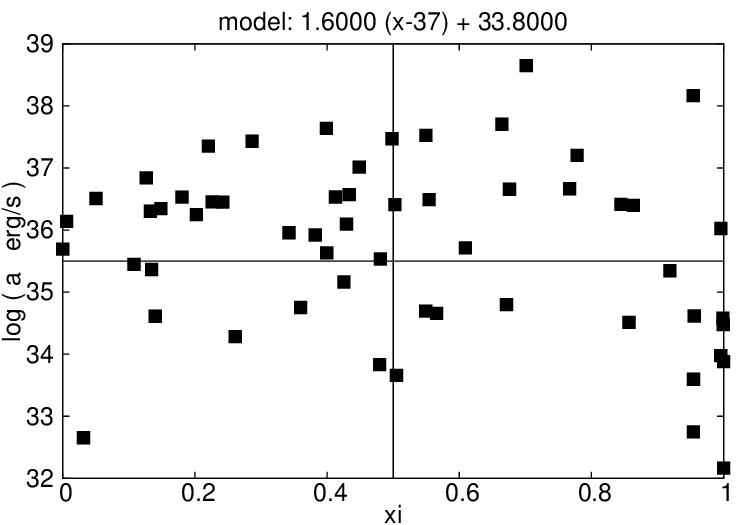} }}
\begin{minipage}{6cm}
\includegraphics[width=58mm]{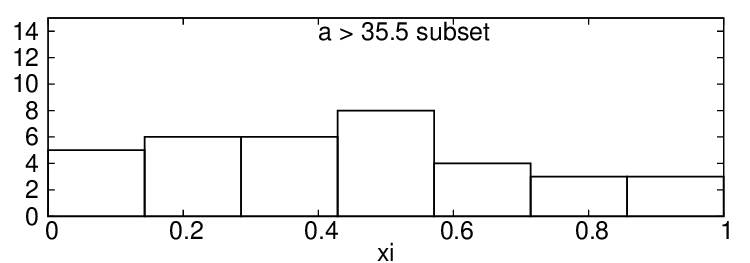} \par
\includegraphics[width=58mm]{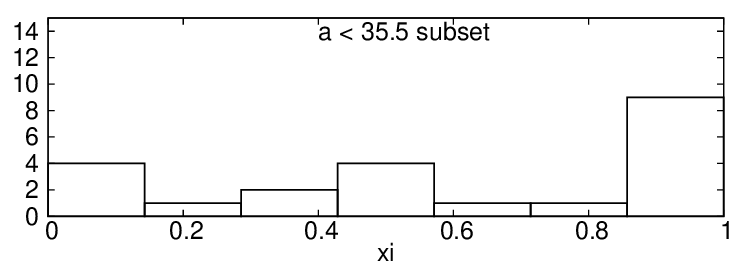}
\end{minipage}

\caption{ \label{rfig7}
The plot of $(\xi_i, a_i)$ for the sampled pulsars (left panels) and
the histogram with respect to $\xi$ (right panels)
for three different model relations,
from the top 
$(c_1, c_2)=$
$(1.0, 33.4)$,
$(0.6, 33.0)$,
and
$(1.6, 33.8)$.
}
\end{center}
\end{figure*}

We need to test the uniformity of $\xi_i$ $\in [0,1)$.
Although this seems straightforward,
it is found that
$\chi^2$ is not sensitive to the slope of the regression line, $c_1$.
The upper left panel of Figure~\ref{rfig7} shows
the scatter plots of $(\xi_i, a_i)$ for the joint Sample SAB,
where the model parameters are
$c_1= 1.0$, $c_2 = 33.4$, $n=2$, ${\cal P}_{\rm mag} =0$,
and $\log F_{lim} = -14.0$.
One can see that 
the distribution with respect to $\xi$ is more or less uniform.
We find  $\chi^2$/dof = 8.4/9.
The middle and bottom plots represent the same plots but for
different slopes, i.e.,
$(c_1, c_2)$,
$=(0.6, 33.0)$,
and
$(1.6, 33.8)$,
respectively.
Although the slopes are significantly different in the two cases,
the distributions with respect to $\xi$ again seem 
more or less uniform
as far as one forgets about distributions with $a$.
We have good values,
$\chi^2/{\rm dof} = 11.2/9$ and
$9.8/9$, respectively.
An important difference is asymmetry in quadrants of the diagram.
In the middle plot (shallow slope), 
a larger population is seen in 
the quadrant with
large $\xi$ and large $a$ 
and in the quadrant with
small $\xi$ and small $a$,
while
in the bottom plot (steep slope),
a larger population is seen in the opposite quadrants.
This tendency is also seen in the histograms (the right column 
of Figure~\ref{rfig7} ) 
made separately for
the two subsets with $a<a_c$ and with $a>a_c$, 
where we take $a_c=35.5$.

A simple $\chi^2$ test for uniformity of $\xi$ is thus found
insensitive to the slope.
This degeneracy can be resolved if we see the distribution
in the $(\xi, a )$ plane.
To have an sensitivity with respect to $c_1$,
we separate the data into two subset,
i.e., a large-$a$ subset and a small-$a$ subset by $a_c =35.5$.
We prepare $N_\xi$ bins for each subsets, and 
the $\chi^2$ test is done for 
$2N_\xi$-bins. 
A suitable number of the bin for the present sample is 
found to be $N_\xi = 7$.


\begin{thebibliography}{99}
\bibitem[Younes et al. (2016)]{GCN 19736} Younes, G., Kouveliotou, C., \& Roberts, O. 2016, GCN, 19736
\bibitem[Kennea et al. (2016)]{GCN 19735} Kennea, J. A., Lien, A. Y., Marshall, F. E., et al. \ 2016, GCN, 19735
\bibitem[Kiziltan et al.(2013)]{2013ApJ...778...66K} Kiziltan, B., Kottas, A., De Yoreo, M., \& Thorsett, S.~E.\ 2013, \apj, 778, 66
\bibitem[Archibald et al.(2016)]{2016arXiv160801007A} Archibald, R.~F., Kaspi, V.~M., Tendulkar, S.~P., \& Scholz, P.\ 2016, arXiv:1608.01007
\bibitem[Abdo et al.(2010)]{2010ApJ...711...64A} Abdo, A.~A., Ackermann, M., Ajello, M., et al.\ 2010, \apj, 711, 64
\bibitem[Abdo et al.(2013)]{2013ApJS..208...17A} Abdo, A.~A., Ajello, M., Allafort, A., et al.\ 2013, \apjs, 208, 17
\bibitem[Acero et al.(2013)]{2013A&A...551A...7A} Acero, F., Gallant, Y., Ballet, J., Renaud, M., \& Terrier, R.\ 2013, \aap, 551, A7
\bibitem[Becker(2009)]{2009ASSL..357...91B} Becker, W.\ 2009, Astrophysics and Space Science Library, 357, 91 
\bibitem[Becker \& Truemper(1997)]{1997A&A...326..682B} Becker, W., \& Truemper, J.\ 1997, \aap, 326, 682
\bibitem[Bejger et al.(2011)]{2011A&A...536A..87B} Bejger, M., Fortin, M., Haensel, P., \& Zdunik, J.~L.\ 2011, \aap, 536, A87
\bibitem[Bogdanov et al.(2014)]{2014ApJ...792L..36B} Bogdanov, S., Ng, C.-Y., \& Kaspi, V.~M.\ 2014, \apjl, 792, L36
\bibitem[Camilo et al.(2009)]{2009ApJ...705....1C} Camilo, F., Ray, P.~S., Ransom, S.~M., et al.\ 2009, \apj, 705, 1
\bibitem[Ciolfi \& Rezzolla(2013)]{2013MNRAS.435L..43C} Ciolfi, R., \& Rezzolla, L.\ 2013, \mnras, 435, L43 
\bibitem[De Luca et al.(2005)]{2005ApJ...623.1051D} De Luca, A., Caraveo, P.~A., Mereghetti, S., Negroni, M., \& Bignami, G.~F.\ 2005, \apj, 623, 1051 
\bibitem[Duncan \& Thompson(1992)]{1992ApJ...392L...9D} Duncan, R.~C., \& Thompson, C.\ 1992, \apjl, 392, L9
\bibitem[Esposito et al.(2010)]{2010MNRAS.405.1787E} Esposito, P., Israel, G.~L., Turolla, R., et al.\ 2010, \mnras, 405, 1787 
\bibitem[Gavriil et al.(2008)]{2008Sci...319.1802G} Gavriil, F.~P., Gonzalez, M.~E., Gotthelf, E.~V., et al.\ 2008, Science, 319, 1802
\bibitem[Gotthelf et al.(2013)]{2013ApJ...765...58G} Gotthelf, E.~V., Halpern, J.~P., \& Alford, J.\ 2013, \apj, 765, 58 . 
\bibitem[Gusakov et al.(2004)]{2004A&A...423.1063G} Gusakov, M.~E., Kaminker, A.~D., Yakovlev, D.~G., \& Gnedin, O.~Y.\ 2004, \aap, 423, 1063
\bibitem[Halpern \& Gotthelf(2010)]{2010ApJ...709..436H} Halpern, J.~P., \& Gotthelf, E.~V.\ 2010, \apj, 709, 436.
\bibitem[He et al.(2013)]{2013ApJ...768...64H} He, C., Ng, C.-Y., \& Kaspi, V.~M.\ 2013, \apj, 768, 64 
\bibitem[Kaaret et al.(2001)]{2001ApJ...546.1159K} Kaaret, P., Marshall, H.~L., Aldcroft, T.~L., et al.\ 2001, \apj, 546, 1159
\bibitem[Kargaltsev \& Pavlov(2008)]{2008AIPC..983..171K} Kargaltsev, O., \& Pavlov, G.~G.\ 2008, 40 Years of Pulsars: Millisecond Pulsars, Magnetars and More, 983, 171
\bibitem[Kargaltsev et al.(2012)]{2012ApJS..201...37K} Kargaltsev, O., Durant, M., Pavlov, G.~G., \& Garmire, G.\ 2012, \apjs, 201, 37
\bibitem[Karpova et al.(2014)]{2014ApJ...789...97K} Karpova, A., Danilenko, A., Shibanov, Y., Shternin, P., \& Zyuzin, D.\ 2014, \apj, 789, 97
\bibitem[Kennea et al.(2002)]{2002astro.ph..2055K} Kennea, J., Cordova, F., Chatterjee, S., et al.\ 2002, arXiv:astro-ph/0202055 
\bibitem[Kuiper \& Hermsen(2015)]{2015MNRAS.449.3827K} Kuiper, L., \& Hermsen, W.\ 2015, \mnras, 449, 3827
\bibitem[Li et al.(2008)]{2008ApJ...682.1166L} Li, X.-H., Lu, F.-J., \& Li, Z.\ 2008, \apj, 682, 1166
\bibitem[Manchester et al.(2005)]{2005AJ....129.1993M} Manchester, R.~N., Hobbs, G.~B., Teoh, A., \& Hobbs, M.\ 2005, \aj, 129, 1993 
\bibitem[Marelli et al.(2011)]{2011ApJ...733...82M} Marelli, M., De Luca, A., \& Caraveo, P.~A.\ 2011, \apj, 733, 82
\bibitem[Marthi et al.(2011)]{2011MNRAS.416.2560M} Marthi, V.~R., Chengalur, J.~N., Gupta, Y., Dewangan, G.~C., \& Bhattacharya, D.\ 2011, \mnras, 416, 2560
\bibitem[McLaughlin et al.(2007)]{2007ApJ...670.1307M} McLaughlin, M.~A., Rea, N., Gaensler, B.~M., et al.\ 2007, \apj, 670, 1307 1
\bibitem[McGowan et al.(2006)]{2006ApJ...639..377M} McGowan, K.~E., Zane, S., Cropper, M., Vestrand, W.~T., \& Ho, C.\ 2006, \apj, 639, 377
\bibitem[Olausen \& Kaspi(2014)]{2014ApJS..212....6O} Olausen, S.~A., \& Kaspi, V.~M.\ 2014, \apjs, 212, 6
\bibitem[Olausen et al.(2013)]{2013ApJ...764....1O} Olausen, S.~A., Zhu, W.~W., Vogel, J.~K., et al.\ 2013, \apj, 764, 1
\bibitem[Ozel et al.(2015)]{2015arXiv151203067O} Ozel, F., Psaltis, D., Arzoumanian, Z., Morsink, S., \& Baubock, M.\ 2015, arXiv:1512.03067
\bibitem[Page et al.(2004)]{2004ApJS..155..623P} Page, D., Lattimer, J.~M., Prakash, M., \& Steiner, A.~W.\ 2004, \apjs, 155, 623 
\bibitem[Pivovaro et al.(1998)]{1998sxmm.confE..83P} Pivovaro, M., Kaspi, V., \& Gotthelf, E.\ 1998, Science with XMM, 83 
\bibitem[Porquet et al.(2003)]{2003A&A...401..197P} Porquet, D., Decourchelle, A., \& Warwick, R.~S.\ 2003, \aap, 401, 197
\bibitem[Posselt et al.(2015)]{2015ApJ...811...96P} Posselt, B., Spence, G., \& Pavlov, G.~G.\ 2015, \apj, 811, 96
\bibitem[Possenti et al.(2002)]{2002A&A...387..993P} Possenti, A., Cerutti, R., Colpi, M., \& Mereghetti, S.\ 2002, \aap, 387, 993
\bibitem[Press et al.(1992)]{press1992} Press, W.H., Flannery, B.P., Teukoisky, S.A., \& Vetterling, 1992, Numerical Recipies in FORTRAN 77, 2nd Eddition (Cambridge, Cambridge University Press)
\bibitem[Prinz \& Becker(2015)]{2015arXiv151107713P} Prinz, T., \& Becker, W.\ 2015, arXiv:1511.07713
\bibitem[Ray et al.(2011)]{2011ApJS..194...17R} Ray, P.~S., Kerr, M., Parent, D., et al.\ 2011, \apjs, 194, 17
\bibitem[Rea et al.(2010)]{2010Sci...330..944R} Rea, N., Esposito, P., Turolla, R., et al.\ 2010, Science, 330, 944 
\bibitem[Rea et al.(2013)]{2013ApJ...770...65R} Rea, N., Israel, G.~L., Pons, J.~A., et al.\ 2013, \apj, 770, 65 
\bibitem[Renaud et al.(2010)]{2010ApJ...716..663R} Renaud, M., Marandon, V., Gotthelf, E.~V., et al.\ 2010, \apj, 716, 663
\bibitem[Safi-Harb \& Kumar(2008)]{2008ApJ...684..532S} Safi-Harb, S., \& Kumar, H.~S.\ 2008, \apj, 684, 532-541
\bibitem[Saito(1998)]{1998.PhD, Univ. of Tokyo} Saito, Y. 1998, Ph.D Thesis, Univ. of Tokyo
\bibitem[Seward \& Wang(1988)]{1988ApJ...332..199S} Seward, F.~D., \& Wang, Z.-R.\ 1988, \apj, 332, 199
\bibitem[Speagle et al.(2011)]{2011ApJ...743..183S} Speagle, J.~S., Kaplan, D.~L., \& van Kerkwijk, M.~H.\ 2011, \apj, 743, 183 
\bibitem[Takata et al.(2011)]{2011MNRAS.415.1827T} Takata, J., Wang, Y., \& Cheng, K.~S.\ 2011, \mnras, 415, 1827
\bibitem[Tepedelenl{\i}o{\v g}lu {\"O}gelman(2005)]{2005ApJ...630L..57T} Tepedelenl{\i}o{\v g}lu, E., {\"O}gelman, H.\ 2005, \apjl, 630, L57
\bibitem[Turolla et al.(2015)]{2015RPPh...78k6901T} Turolla, R., Zane, S., \& Watts, A.~L.\ 2015, Reports on Progress in Physics, 78, 116901 
\bibitem[Vigan{\`o} et al.(2013)]{2013MNRAS.434..123V} Vigan{\`o}, D., Rea, N., Pons, J.~A., et al.\ 2013, \mnras, 434, 123
\bibitem[Vink et al.(2011)]{2011ApJ...727..131V} Vink, J., Bamba, A., \& Yamazaki, R.\ 2011, \apj, 727, 131
\bibitem[Yakovlev et al.(2011)]{2011MNRAS.411.1977Y} Yakovlev, D.~G., Ho, W.~C.~G., Shternin, P.~S., Heinke, C.~O., \& Potekhin, A.~Y.\ 2011, \mnras, 411, 1977
\bibitem[Zhu et al.(2009)]{2009ApJ...704.1321Z} Zhu, W., Kaspi, V.~M., Gonzalez, M.~E., \& Lyne, A.~G.\ 2009, \apj, 704, 1321
\bibitem[Zhu et al.(2011)]{2011ApJ...734...44Z} Zhu, W.~W., Kaspi, V.~M., McLaughlin, M.~A., et al.\ 2011, \apj, 734, 44
\end{thebibliography}
\end{document}